\documentclass[pdflatex,sn-nature]{sn-jnl}

\usepackage{graphicx}%
\graphicspath{ {./images/} }
\usepackage{multirow}%
\usepackage{amsmath,amssymb,amsfonts}%
\usepackage{amsthm}%
\usepackage{mathrsfs}%
\usepackage[title]{appendix}%
\usepackage{xcolor}%
\usepackage{textcomp}%
\usepackage{manyfoot}%
\usepackage{booktabs}%
\usepackage{algorithm}%
\usepackage{algpseudocode}%
\usepackage{listings}%
\usepackage{hyperref}       
\usepackage{url}            
\usepackage{nicefrac}       
\usepackage{microtype}      
\usepackage{lipsum}
\usepackage{float}
\usepackage{enumitem}
\usepackage{siunitx}
\usepackage{caption}
\usepackage{makecell} 
\usepackage{rotating}
\begin{document}

\title{Fusion of multi-source precipitation records via coordinate-based generative models}


\author[1]{\fnm{Sencan} \sur{Sun}}

\author[2]{\fnm{Congyi} \sur{Nai}}

\author*[2]{\fnm{Baoxiang} \sur{Pan}}\email{panbaoxiang@lasg.iap.ac.cn}

\author[3]{\fnm{Wentao} \sur{Li}}

\author[4]{\fnm{Lu} \sur{Li}}

\author[5]{\fnm{Xin} \sur{Li}}

\author[6]{\fnm{Efi} \sur{Foufoula-Georgiou}}

\author*[1]{\fnm{Yanluan} \sur{Lin}}\email{yanluan@tsinghua.edu.cn}

\affil*[1]{\orgdiv{Ministry of Education Key Laboratory for Earth System Modeling, Department of Earth System Science}, \orgname{Tsinghua University}, \orgaddress{\city{Beijing}, \postcode{100084}, \country{China}}}

\affil*[2]{\orgdiv{Institute of Atmospheric Physics}, \orgname{Chinese Academy of Sciences}, \orgaddress{\city{Beijing}, \country{China}}}

\affil[3]{\orgdiv{State Key Laboratory of Hydrology–Water Resources and Hydraulic Engineering and College of Hydrology \& Water Resources}, \orgname{Hohai University}, \orgaddress{\city{Nanjing}, \country{China}}}

\affil[4]{\orgdiv{School of Atmospheric Sciences}, \orgname{Sun Yat-sen University}, \orgaddress{\city{Guangzhou}, \country{China}}}

\affil[5]{\orgdiv{Institute of Tibetan Plateau Research}, \orgname{Chinese Academy of
Sciences}, \orgaddress{\city{Beijing}, \country{China}}}

\affil[6]{\orgdiv{Department of Earth System Science}, \orgname{University of California Irvine}, \orgaddress{\city{Irvine}, \state{CA}, \country{USA}}}

%
%
%
%
%

\abstract{

Precipitation remains one of the most challenging climate variables to observe and predict accurately. Existing datasets face intricate trade-offs: gauge observations are relatively trustworthy but sparse, satellites provide global coverage with retrieval uncertainties, and numerical models offer physical consistency but are biased and computationally intensive. Here we introduce PRIMER (Precipitation Record Infinite MERging), a deep generative framework that fuses these complementary sources to produce accurate, high-resolution, full-coverage precipitation estimates.
PRIMER employs a coordinate-based diffusion model that learns from arbitrary spatial locations and associated precipitation values, enabling seamless integration of gridded data and irregular gauge observations. Through two-stage training—first learning large-scale patterns, then refining with accurate gauge measurements—PRIMER captures both large-scale climatology and local precision. Once trained, it can downscale forecasts, interpolate sparse observations, and correct systematic biases within a principled Bayesian framework.
Using gauge observations as ground truth, PRIMER effectively corrects biases in existing datasets, yielding statistically significant error reductions at most stations and furthermore enhancing the spatial coherence of precipitation fields. Crucially, it generalizes without retraining, correcting biases in operational forecasts it has never seen. This demonstrates how generative AI can transform Earth system science by combining imperfect data, providing a scalable solution for global precipitation monitoring and prediction.

}


\maketitle

\section{Introduction}

Precipitation--when, where, and how much water falls from the sky to the earth surface--governs freshwater availability, agricultural productivity, flood hazards, and ecosystem health across the globe~\cite{kotz2022effect}.
Despite its significance, precipitation remains one of the most challenging climate variables to observe and predict accurately.
This challenge stems from precipitation's fundamental nature: unlike most climate variables that vary smoothly across space and time, precipitation manifests as discrete, intermittent pulses with striking discontinuities~\cite{sun2006often, pendergrass2018uneven}.
These processes depend crucially on small-scale cloud microphysics processes~\cite{stevens2009untangling} that remain poorly understood or simulated.
Besides, these processes are highly sensitive to environmental conditions: small perturbations in temperature, humidity, or aerosol concentrations can determine whether clouds produce no rain, light drizzle, or torrential downpours~\cite{birch2013impact, prein2023multi}. Furthermore, the triggering and organization of convection--the primary mechanism for intense precipitation--depends on complex interactions between boundary layer turbulence~\cite{teixeira2008parameterization}, atmospheric stability~\cite{lepore2015temperature}, and mesoscale circulations~\cite{arakawa2004cumulus, houze2004mesoscale} that remain computationally prohibitive to simulate explicitly. 
These complexities create fundamental observational and predictive challenges.



Currently, we rely upon three sources to derive precipitation information, namely in-situ gauge observations, remote sensing, and numerical simulation that potentially assimilate in-situ and remote-sensed data~\cite{sun2018review}. 
Each of these three sources comes with inherent limitations regarding their accuracy, coverage, and resolution.
Ground-based observations from rain gauges provide the most direct and accurate measurements at point locations. However, gauge networks exhibit severe spatial limitations: even 2.5$^\circ\,{\times}\,$2.5$^\circ$ grid cells contain less than two gauges on avearge~\cite{Global_precipitation_measurement}, let alone the oceanic and remote regions.
Satellite remote sensing offers near-global coverage, but measures precipitation indirectly. Passive microwave sensors on polar-orbiting satellites detect emission and scattering signatures from hydrometeors, providing relatively direct estimates but with limited temporal sampling (2-4 observations per day)~\cite{hou2014global}. Infrared sensors on geostationary satellites offer frequent observations (every 10-30 minutes) but only measure cloud-top temperatures, requiring empirical relationships to infer surface precipitation--a particularly poor assumption for shallow, warm clouds that produce significant precipitation in tropical and maritime regions~\cite{levizzani2002review}. Numerical weather prediction and reanalysis products provide physically consistent, complete spatiotemporal coverage by assimilating available observations into dynamical models~\cite{bauer2015quiet}. However, precipitation in these systems emerges as the end result of a complex chain of parameterized processes—radiation, convection, cloud microphysics, and boundary layer turbulence—each contributing its own errors~\cite{tapiador2019precipitation}, with their errors compounding multiplicatively. The consequence of these observational and simulational limitations is profound: current precipitation datasets often disagree by as much as the signal itself~\cite{sun2018review, tapiador2019precipitation}. In tropical regions, the spread among different products can exceed 300 mm/hr of the mean precipitation~\cite{sun2006often}, fundamentally limiting our ability to close the global water budget, validate climate models, or provide reliable information for water resource management.

A promising solution to these challenges lies in data fusion--leveraging the complementary strengths of multiple data sources to produce precipitation estimates that surpass any individual source in accuracy, resolution, and coverage~\cite{he2020first, ma2018performance, baez2020rf, ur2019appraisal, yumnam2022quantile, xie2011conceptual, woldemeskel2013merging, fan2021comparative, zhang2021merging, bhuiyan2020machine, bhuiyan2018nonparametric, wu2020spatiotemporal, shen2014high}. Among data fusion approaches, Bayesian methods provide the most principled solution. The key insight is elegant: by deriving an informative prior distribution from all available sources, we can encode existing knowledge in a statistically coherent form. Once established, this prior can be updated via Bayes' theorem with any new observation--accounting for each source's unique error characteristics through tailored likelihood functions~\cite{box2011bayesian, wu2023bayesian}. The framework naturally weights observations by their reliability and propagates uncertainties to yield full posterior distributions~\cite{price2023gencast}, essential for risk assessment.

Recent advances in deep generative models~\cite{goodfellow2020generative, kingma2013auto, papamakarios2021normalizing}, particularly probabilistic diffusion models~\cite{ho2020denoising, song2022denoisingdiffusionimplicitmodels}, offer a transformative opportunity for implementing the above Bayesian framework. Diffusion models have demonstrated remarkable ability to learn complex, high-dimensional distributions--from natural images~\cite{dhariwal2021diffusionmodelsbeatgans} to protein structures~\cite{yim2024diffusion}--making them ideal candidates for capturing the intricate patterns of precipitation. Crucially, these models can serve as learned priors for Bayesian inference, where their probabilistic foundations enable principled uncertainty quantification. Once trained, they function as ``plug-and-play'' priors~\cite{daras2024surveydiffusionmodelsinverse, zheng2025inversebenchbenchmarkingplugandplaydiffusion, hess2024fast}: the same learned distribution can be applied to diverse inference tasks--bias correction, downscaling, or gap-filling--by simply changing the likelihood function without retraining. 
Despite the promises, implementing this framework for precipitation faces three fundamental challenges. First, precipitation's extreme spatiotemporal variability--from localized convective cells to continental-scale fronts--makes it extraordinarily difficult to be captured in a single prior distribution. Second, constructing an informative prior becomes paradoxical when no individual data source is trustworthy or comprehensive. Each source captures different aspects of precipitation across mismatched scales, creating a circular dependency where we need accurate data to build a prior, yet need a prior to evaluate data accuracy. 
Third, even with a reasonable prior, posterior sampling is challenging due to the high dimensionality of precipitation fields and the complexity of observation likelihoods.
These barriers define the frontier for deploying generative AI in Earth system science, demanding innovations that transcend conventional generative modelling approaches.


To address these challenges, we introduce PRIMER (Precipitation Record Infinite MERging), a novel framework that reconceptualizes how diffusion models can learn from imperfect, heterogeneous precipitation records here after for relevant probabilistic inference tasks. 
Our key insight is that probabilistic diffusion models need not be trained on perfect samples -- instead, they can be viewed as spectral regression models that progressively learn from low-frequency structures to high-frequency details as we gradually corrupt the target distribution using Gaussain noise~\cite{dieleman2024spectral}. This property enables us to construct an informative prior by learning conditional distributions of precipitation patterns for each data source, where the conditioning explicitly captures each dataset's characteristic biases.

\begin{figure}[htbp]
    \centering
    \includegraphics[width=1\textwidth,page=1]{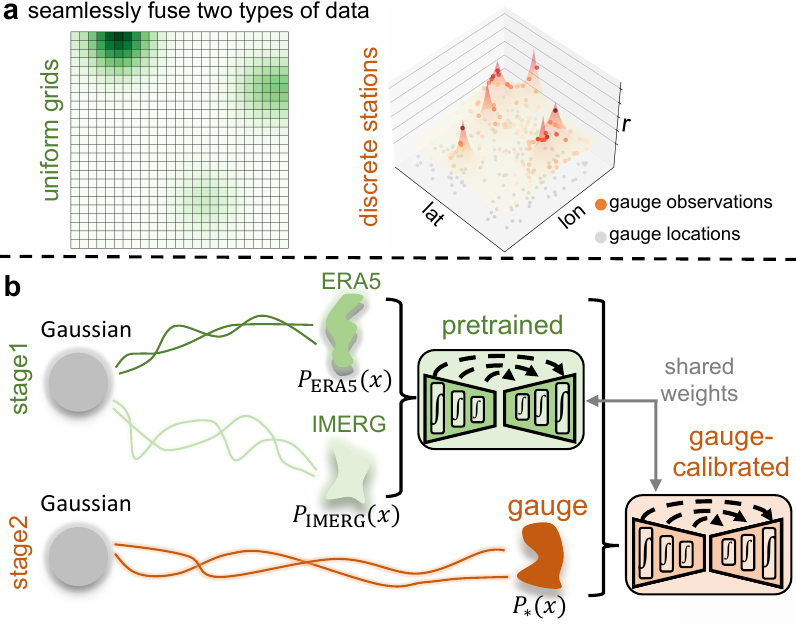} 
\caption{
\textbf{Overview of PRIMER.}
(\textbf{a}) No single precipitation dataset provides uniformly reliable estimates across all spatial scales. PRIMER addresses this challenge by integrating heterogeneous data sources, including gridded reanalysis (e.g., ERA5), satellite-retrieved products (e.g., IMERG), and sparse but accurate in-situ gauge observations. 
(\textbf{b}) Our goal is to fuse information from these diverse datasets into a coherent and accurate prior distribution. PRIMER is trained in two stages. In \textbf{Stage 1}, the model is pretrained on gridded datasets to learn baseline priors $P_{\mathrm{ERA5}}(x)$ and $P_{\mathrm{IMERG}}(x)$. In \textbf{Stage 2}, it is fine-tuned using gauge observations and their corresponding locations to produce a updated prior $P_{*}(x)$. Weight sharing across data sources enables the model to leverage large-scale spatial patterns from gridded products while incorporating localized constraints from sparse gauge observations. 
In the following experiments, we will demonstrate that $P_{*}(x)$ yields superior accuracy compared to $P_{\mathrm{ERA5}}(x)$ and $P_{\mathrm{IMERG}}(x)$.
}
    \label{fig:Fig1}
\end{figure}

Implementing this multi-source learning faces a fundamental architectural barrier. Conventionally, diffusion models work on samples residing on fixed-resolution grids~\cite{ronneberger2015unetconvolutionalnetworksbiomedical}, forcing us to interpolate heterogeneous observations to common resolutions. This interpolation is particularly destructive for precipitation: it smooths sharp gradients at convective boundaries, introduces artificial correlations between sparse gauge points, and—most critically—destroys the very precision that makes gauges valuable. For sparse gauge networks covering less than 1\% of the domain, interpolation essentially fabricates information that doesn't exist. We therefore require an architecture that can learn priors directly from each source's native sampling structure.
This necessity drives our adoption of coordinate-based diffusion models, which represent precipitation as continuous spatial fields $x : \mathbb{R}^2\rightarrow \mathbb{R}$ rather than tensors. In this formulation, both dense grids and sparse gauge observations are simply different sampling patterns of the same underlying field. PRIMER directly learns from arbitrarily and sparsely distributed points—each defined by its latitude, longitude, and precipitation intensity—without relying on spatial interpolation~(see Fig.\ref{fig:Fig1}a)—gauge observations influence the function locally while gridded data constrain large-scale structure. 
Our two-stage training strategy is thus a natural choice: we first learn the baseline priors \( P_{\mathrm{ERA5}}(x) \) and \( P_{\mathrm{IMERG}}(x) \), which represent the climatological distributions of precipitation fields \( x \) derived from climate reanalysis, i.e., fifth generation ECMWF atmospheric reanalysis (ERA5), and satellite-based retrieval dataset, i.e., Integrated Multi-satellitE Retrievals for GPM (IMERG).
We then fine-tune the model using gauge observations to incorporate local accuracy, yielding the updated prior \( P_{*}(x) \) (Fig.~\ref{fig:Fig1}b; star indicates that it supposes to be a better prior).
The coordinate-based representation ensures that gauge information enhances rather than corrupts the prior, as each source contributes at its natural scale.
Once trained, PRIMER supports diverse applications through principled posterior sampling: given observations $\mathcal{O}$—whether from biased satellites, sparse gauges, or coarse forecasts—we can sample from posterior $P_{*}(x \mid \mathcal{O})$ to produce improved ensemble estimates. Empirical evaluations demonstrate the effectiveness of our approach: when assessed against approximately 1,000 independent rain gauges across a diverse set of precipitation events, PRIMER reduces errors at the majority of sites. It also enhances the representation of extreme events and improves the realism of spatial structures, and generalizes effectively to previously unseen operational forecasts without retraining. By transforming the challenge of heterogeneous, imperfect data from a limitation into a strength, PRIMER establishes a new paradigm for precipitation data fusion that extends naturally to other Earth system variables plagued by observational trade-offs.

\section{Results}

\subsection{Reproducing climatological distributions}

\begin{figure}[htbp]
    \centering
    \includegraphics[width=0.9\textwidth,page=1]{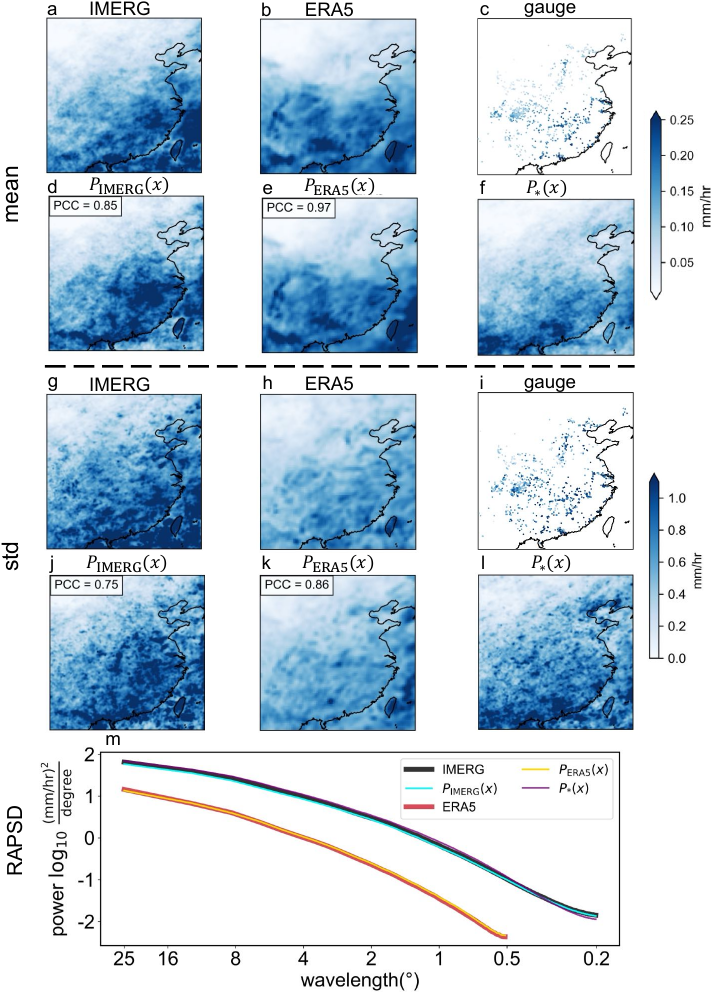} 
    \caption{\textbf{Climatological consistency between learned priors and reference datasets.} 
        \textbf{a–f}, Spatial distributions of mean precipitation from IMERG (\textbf{a}), ERA5 (\textbf{b}), gauge observations (\textbf{c}), the learned prior $P_{\mathrm{IMERG}}(x)$ (\textbf{d}), $P_{\mathrm{ERA5}}(x)$ (\textbf{e}), and the final updated prior $P_{*}(x)$ (\textbf{f}). 
        \textbf{g–l}, Standard deviation fields analogous to \textbf{a–f}. 
        Pearson correlation coefficients (PCCs) between each learned prior and its corresponding reference (IMERG or ERA5) are indicated in the upper-left corner of relevant panels.
        \textbf{m}, Radially averaged power spectral density (RAPSD) as a function of spatial wavelength (in degrees), quantifying the spatial structure of precipitation fields. The learned priors $P_{\mathrm{IMERG}}(x)$ and $P_{\mathrm{ERA5}}(x)$ closely follow their references, and $P_{*}(x)$ captures consistent multiscale characteristics.
        All statistics are computed from 1,000 randomly sampled realizations. Colorbars represent the units for each corresponding row.
    }
	\label{fig:unconditional_generation}
\end{figure}

The gist of the PRIMER methodology is to learn a trustworthy prior distribution of precipitation fields, thereafter applying it for a broad range of relevant probabilistic inference tasks, so as for accurate, high-resolution, full-coverage precipitation estimates and forecasts. Before verifying the probabilistic inference results, we should ensure the accuracy of the learned prior distribution. 
	As directly evaluating such high-dimensional
	priors is intractable, we instead assess their  statistical properties as proxies~\cite{klein2013climate, zhang2022e3sm, lee2024systematic}.
We compare unconditionally generated samples from $P_{\mathrm{IMERG}}(x)$, $P_{\mathrm{ERA5}}(x)$, and the final prior $P_{*}(x)$ against their respective reference datasets.
In particular, we focus on the climatological mean and standard deviation of precipitation (Fig.~\ref{fig:unconditional_generation}).
At the grid-point level, the agreement is clear. For mean precipitation (Fig.~\ref{fig:unconditional_generation}a–f), both $P_{\mathrm{IMERG}}(x)$ and $P_{\mathrm{ERA5}}(x)$ exhibit strong spatial correspondence with IMERG and ERA5, achieving Pearson correlation coefficients (PCCs) of 0.85 and 0.97, respectively. The standard deviation fields (Fig.~\ref{fig:unconditional_generation}g–l) are likewise well reproduced (PCC = 0.75 and 0.86), highlighting PRIMER’s capacity to represent not just the average precipitation distribution but also its variance.
Notably, we also introduce the updated prior $P_{*}(x)$, constructed by fine-tuning PRIMER using sparse but reliable gauge observations (data description is available in Method~\ref{data_description}). Despite the limited spatial coverage of gauge observations, this calibration yields a climatologically consistent prior that preserves spatial structures learned from the gridded products while injecting localized realism. This “climatological jailbreak” illustrates how PRIMER can adapt to sparse in situ records without compromising coherence across scales.
To further evaluate spatial structure, we perform a radially averaged power spectral density (RAPSD) analysis (Fig.~\ref{fig:unconditional_generation}m), which confirms that the learned priors accurately recover the multiscale spectral characteristics of the reference datasets, especially across mesoscale wavelengths, which are crucial for convective processes (see also Supplementary Information (SI) Fig.~\ref{SI:SI_pdf_RAPSD}).
Additional statistical evaluations—including precipitation frequency, extremes, skewness, and Empirical Orthogonal Function (EOF) modes—are provided in the SI Fig.~\ref{SI:SI_eof_skew}.

\subsection{Case study on high-impact events}
\label{Case study on extreme events}

\begin{figure}[htbp]
\centering
\includegraphics[width=\textwidth,page=1]{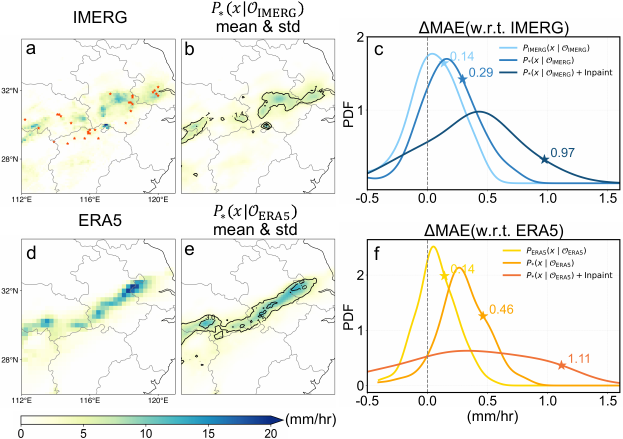}

\caption{\textbf{Case study of a Meiyu precipitation event on 2 July 2016 at 05 UTC.} 
\textbf{a}, Precipitation field from IMERG at the target time, with gauge locations shown as red dots (used as ground truth for evaluation).
\textbf{b}, Posterior mean and standard deviation from $P_{*}(x \mid \mathcal{O}_{\mathrm{IMERG}})$ inferred by PRIMER. 
\textbf{c}, Probability density functions (PDFs) of changes in mean absolute error ($\Delta$MAE), computed at gauge locations by comparing 100 posterior samples and the original IMERG data against observations. For each posterior sample, $\Delta$MAE is calculated as the difference between the sample's MAE and that of IMERG, with positive values indicating effective bias correction by PRIMER.
\textbf{d}, Precipitation field from ERA5. 
\textbf{e}, Posterior mean and standard deviation from $P_{*}(x \mid \mathcal{O}_{\mathrm{ERA5}})$. 
\textbf{f}, PDFs of $\Delta$MAE relative to ERA5, analogous to \textbf{c}. 
In \textbf{c,f}, different curves represent various posterior distributions as labeled; ensemble means are marked with stars.
}

\label{fig:meiyu_hubei}
\end{figure}

The previous section evaluated PRIMER’s ability to match climatological distributions. After Stage 2 fine-tuning, the updated prior \(P_{*}(x)\) is expected to align more closely with gauge observations; however, its actual skill remains to be validated through posterior sampling experiments.
To this end, we perform posterior sampling using different priors while conditioning on the same observations \(\mathcal{O}\). By comparing the posterior samples against the held-out gauge data, we directly assess the impact of the prior on posterior accuracy, thereby quantifying how much fine-tuning improves alignment with real-world observations.
We examine three representative high-impact events. These events were selected to span a wide range of precipitation regimes, including prolonged precipitation associated with the Meiyu front, heavy precipitation driven by landfalling typhoons, and localized convective extremes.
The primary case, which occurred over Hubei Province, China, during the East Asian summer monsoon on 2 July 2016, is shown in Fig.~\ref{fig:meiyu_hubei}; additional examples are provided in Fig.~\ref{SI:SI_Meranti} and Fig.~\ref{SI:SI_720event}. 

To evaluate the effectiveness of the posterior sampling, we define a relative skill metric, $\Delta \mathcal{M}$, based on standard performance scores, including mean absolute error (MAE) and the continuous ranked probability score (CRPS). The CRPS provides a probabilistic measure of an ensemble system’s accuracy (see Method~\ref{CRPS}). For each metric, $\Delta \mathcal{M}$ quantifies the improvement relative to the original precipitation datasets (ERA5 or IMERG), with positive values indicating reduced error or enhanced skill. All evaluations are conducted at a spatial resolution of 0.1°, where ERA5, IMERG, and posterior samples are compared against independent gauge observations treated as ground truth.

As shown in Fig.~\ref{fig:meiyu_hubei}c and Fig.~\ref{fig:meiyu_hubei}f, the updated prior \(P_{*}(x)\) substantially outperforms baseline priors derived from reanalysis (ERA5) and satellite retrievals (IMERG). The ensemble-mean $\Delta$MAE decreases from 0.46~mm/hr for \(P_{*}(x \mid \mathcal{O}_{\mathrm{ERA5}})\) to 0.14~mm/hr for \(P_{\mathrm{ERA5}}(x \mid \mathcal{O}_{\mathrm{ERA5}})\); a similar improvement is observed in the IMERG case, where the $\Delta$MAE decreases from 0.29~mm/hr to 0.14~mm/hr. These gains extend beyond ensemble means: across individual samples, $\Delta$MAE values for $P_{\mathrm{ERA5}}(x \mid \mathcal{O}_{\mathrm{ERA5}})$ are consistently lower than those for $P_{*}(x \mid \mathcal{O}_{\mathrm{ERA5}})$ (see SI Fig.~\ref{SI:SI_meiyuhubei}).   
PRIMER allows the posterior sampling process to incorporate not only the background field but also additional gauge observations, if available. To evaluate this capability, we conduct an experiment where a subset (20\%) of gauge observations are assimilated during sampling (denoted as “+ Inpaint”). This additional constraint significantly enhances accuracy, with posterior mean $\Delta$MAE increasing to 1.11~mm/hr and 0.97~mm/hr for the ERA5 and IMERG cases, respectively. This highlights PRIMER’s capacity to integrate background field with observational data.
Spectral analysis further highlights distinctions among posterior samples (see SI Fig.~\ref{SI:SI_meiyuhubei}). While $P_{\mathrm{ERA5}}(x \mid \mathcal{O}_{\mathrm{ERA5}})$ retains low-frequency biases, both $P_{*}(x \mid \mathcal{O}_{\mathrm{ERA5}})$ and its \textrm{Inpaint} variant enhance high-frequency components.

\subsection{Statistical verifications}

We applied PRIMER to a curated test set of 150 precipitation events from 2016, selected based on the criteria detailed in SI~\ref{Selection of test precipitation events in 2016}. 
For each event, 50 posterior samples were drawn from $P_{*}(x \mid \mathcal{O})$, where $\mathcal{O}$ corresponds to raw data from either ERA5 or IMERG.
In this process, PRIMER downscales ERA5 data to 0.1° resolution and performs bias correction, while directly correcting biases in IMERG. To evaluate the improved accuracy of the prior \(P_{*}(x)\), we also conducted posterior sampling using the baseline priors (\(P_{\mathrm{ERA5}}(x)\) and \(P_{\mathrm{IMERG}}(x)\)) under identical settings. 
At each gauge location, we computed the mean absolute error (MAE) and continuous ranked probability score (CRPS) of the posterior distributions. MAE was calculated using the ensemble mean of each posterior compared against the corresponding gauge observation, while CRPS assessed the full probabilistic accuracy. We then calculated differences in both metrics between the baseline posteriors—\(P_{\mathrm{ERA5}}(x \mid \mathcal{O})\) and \(P_{\mathrm{IMERG}}(x \mid \mathcal{O})\)—and the posterior \(P_{*}(x \mid \mathcal{O})\). Specifically, \(\Delta\mathrm{MAE}\) and \(\Delta\mathrm{CRPS}\) are defined as the baseline scores minus those of \(P_{*}(x \mid \mathcal{O})\), such that positive values indicate improved performance.

\begin{figure}[htbp]
\centering
\includegraphics[width=1\textwidth]{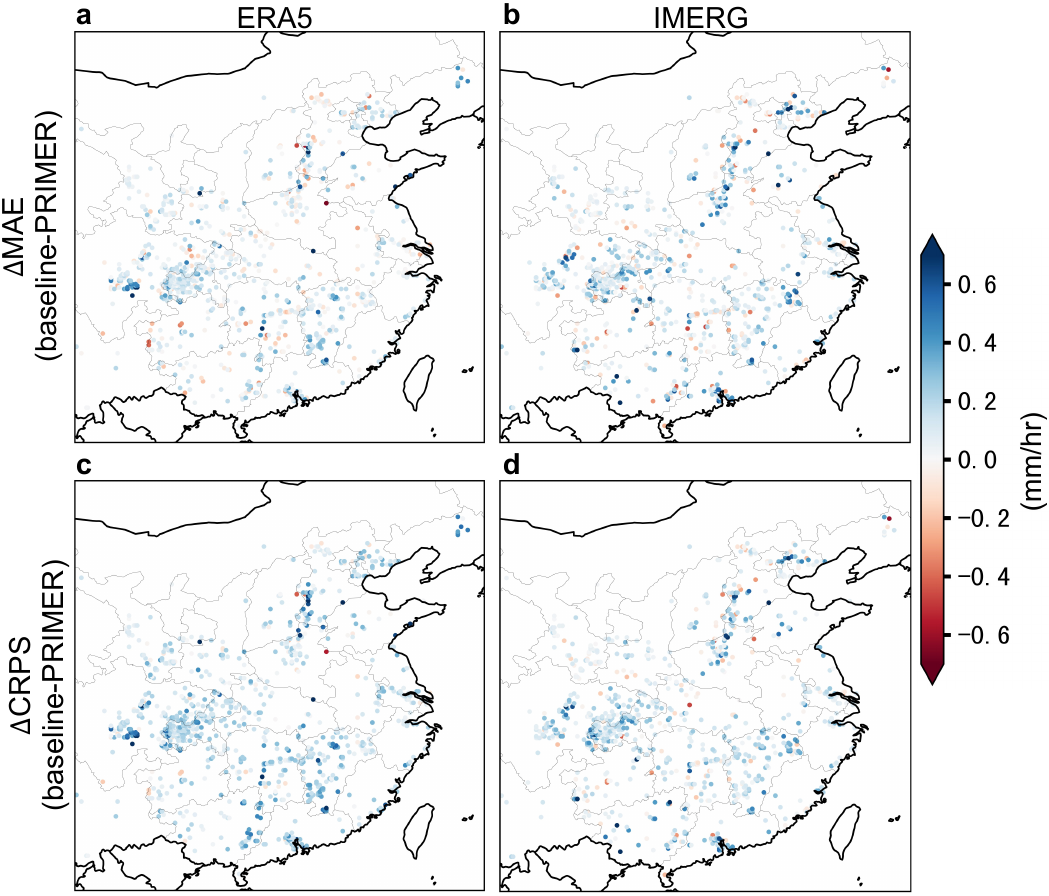}

\caption{\textbf{Improvement of PRIMER over the baseline in bias correction of existing precipitation datasets.} 
\textbf{a,b}, Improvements of PRIMER over the baseline after bias correction of ERA5 (\textbf{a}) and IMERG (\textbf{b}) evaluated by mean absolute error (MAE).  
\textbf{c,d}, Corresponding changes in continuous ranked probability score (CRPS) under the same settings.  
Each dot denotes a gauge station, with errors evaluated against gauge observations (serving as ground truth). 
Positive values (blue) indicate improved performance of PRIMER relative to the baseline model, while negative values (red) denote deterioration. 
The predominance of positive values suggests that PRIMER consistently achieves better bias correction effect, likely due to its ability to learn a more accurate prior distribution $P_{*}(x)$ by leveraging sparse, discrete gauge observations. 
Error statistics are based on 150 precipitation events from 2016. 
For spatial distributions of each model’s MAE relative to ERA5 or IMERG, refer to SI Fig.~\ref{SI:SI_station_test}.
}

\label{fig:station_test_baseline_minus_PRIMER}
\end{figure}

Figures~\ref{fig:station_test_baseline_minus_PRIMER}a–b reveal widespread reductions in mean absolute error (MAE), highlighting PRIMER’s ability to systematically correct biases inherent in the original datasets, outperforming the baseline priors \(P_{\mathrm{ERA5}}(x)\) and \(P_{\mathrm{IMERG}}(x)\). Figures~\ref{fig:station_test_baseline_minus_PRIMER}c–d show even deeper blue tones in CRPS, indicating more substantial improvements in probabilistic estimates. This suggests that PRIMER not only improves point estimates but also models the full posterior distribution more accurately, thereby reducing uncertainty and enhancing the reliability of ensemble-mean outputs. Notably, the largest improvements are observed in the Sichuan Basin and Pearl River Delta—regions with dense populations and strong economic activity—likely due to the higher density of gauge observations available for Stage 2 fine-tuning.

\begin{figure}
\centering
\includegraphics[width=1.02\textwidth]{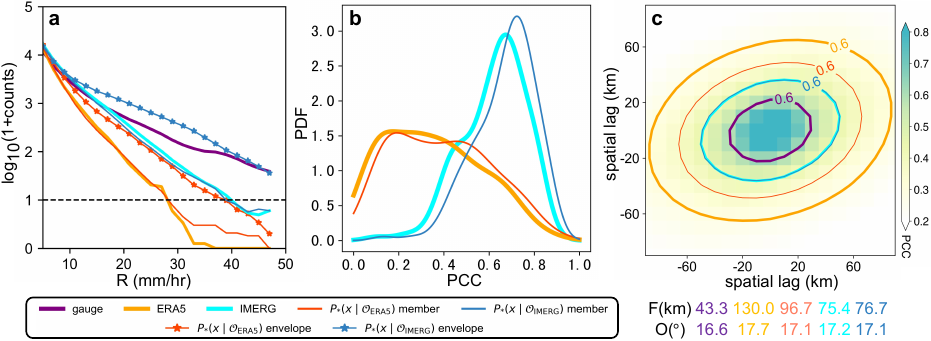}

\caption{\textbf{Improved intensity distribution and spatial coherence after bias correction of existing datasets.}
\textbf{a}, Log-transformed histogram of precipitation intensity (2 mm/hr bins) at only gauge locations, aggregated over test sets. This panel highlights the ability of different datasets to reproduce the tail of the precipitation distribution (with purple line as the ground truth). 
\textbf{b}, Probability density functions (PDFs) of pixel-wise Pearson correlation coefficients (PCCs) between each dataset and the individual gauge observations. Higher PCC values indicate better structural fidelity to ground truth. 
\textbf{c}, Spatial lag correlation maps, with the 0.6 PCC contour visualized for each dataset. Elliptical fits to these contours are used to quantify the spatial coherence, including the major axis length (focal distance, $F$) and orientation angle ($O$), as summarized below \textbf{c}. Colors in panels \textbf{a}–\textbf{c} are illustrated in the below legend.}

\label{fig:gain}
\end{figure}

Beyond reducing pointwise error, PRIMER also enhances the physical realism of existing precipitation datasets. 
To comprehensively evaluate the performance of PRIMER, we adopt two complementary perspectives: the \textit{member} view and the \textit{envelope} view. The \textit{member} view analyzes statistics from a single sample, representing one physically plausible realization. In contrast, the \textit{envelope} is constructed by selecting, at each gauge location, for a given event, the maximum precipitation value across 50 posterior samples. 
As illustrated in Fig.~\ref{fig:gain}a, both $P_{*}(x \mid \mathcal{O}_{\mathrm{ERA5}})$ and $P_{*}(x \mid \mathcal{O}_{\mathrm{IMERG}})$ more accurately reproduce the frequency distribution of precipitation, particularly at higher intensities.
Both perspectives reveal improvements in the representation of heavy precipitation tails compared to the existing datasets, underscoring PRIMER’s capacity to detect high-impact precipitation events that are often underrepresented in deterministic products. Improvements in spatial structure are further quantified using pixel-wise Pearson correlation coefficients (PCCs) with respect to gauge observations (Fig.~\ref{fig:gain}b). $P_{*}(x \mid \mathcal{O}_{\mathrm{ERA5}})$ and $P_{*}(x \mid \mathcal{O}_{\mathrm{IMERG}})$ show markedly enhanced structural agreement relative to existing datasets, suggesting that PRIMER not only reduces local biases but also restores spatial coherence. While various methods have been proposed to assess spatial organization and feature propagation~\cite{guilloteau2021well, guilloteau2022well}, we employ a simplified yet informative diagnostic based on two-dimensional spatial lagged correlation coefficient (Method~\ref{spatial_autocorrelation}, Fig.~\ref{fig:gain}c). Physically, these correlation characterizes how anomalies at a reference point are spatially linked to those at surrounding locations, thereby revealing key features of precipitation system organization. We approximate the 0.6 correlation contour with an ellipse and extract two geometric descriptors: the focal length ($F$), indicative of spatial extent, and the orientation ($O$), which captures the dominant directional alignment. 
Results show that both $P_{*}(x \mid \mathcal{O}_{\mathrm{ERA5}})$ and $P_{*}(x \mid \mathcal{O}_{\mathrm{IMERG}})$ produce orientations that are more consistent with reference orientations derived from gauge observations, indicating improved spatial alignment. In terms of focal length, $P_{*}(x \mid \mathcal{O}_{\mathrm{ERA5}})$ exhibits a clear reduction, while $P_{*}(x \mid \mathcal{O}_{\mathrm{IMERG}})$ shows no substantial improvement. These results demonstrate PRIMER’s effectiveness in correcting spatial anisotropy of precipitation systems.

\subsection{Improving operational forecasts without retraining}

\begin{figure}[htbp]
\centering
\includegraphics[width=1\textwidth]{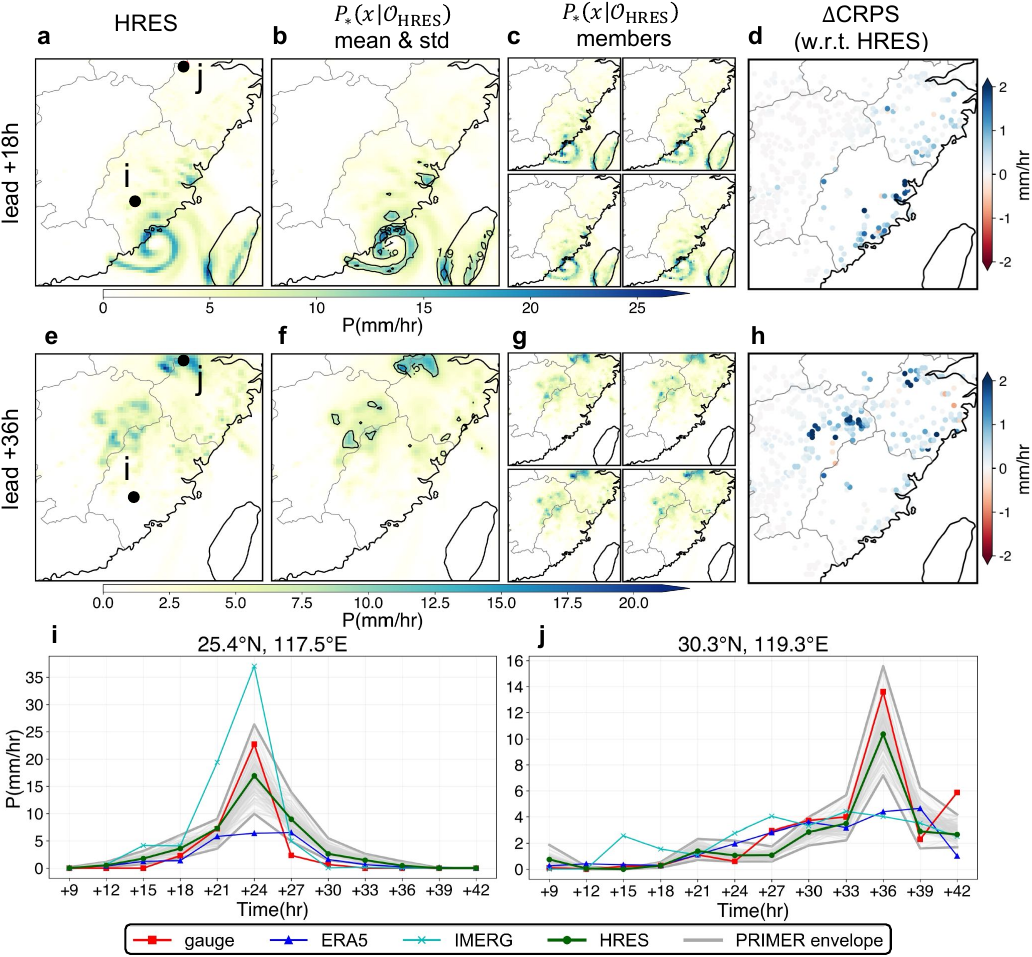}
\caption{\textbf{Bias-correction for operational forecasts without retraining.} 
\textbf{a}, \textbf{e}: HRES forecasts at 18-hr and 36-hr lead times (other lead times see Fig.~\ref{SI:SI_HRES}), initialized at 00:00 UTC on 14 September 2016, coinciding with the landfall of Typhoon Meranti.
\textbf{b}, \textbf{f}: Ensemble means. 
\textbf{c}, \textbf{g}: Four representative ensemble members, illustrating internal variability and structural diversity. 
\textbf{d}, \textbf{h}: Spatial distribution of $\Delta \mathrm{CRPS}$, with blue indicating improvement and red indicating deterioration. 
\textbf{i}, \textbf{j}: Precipitation time series at two representative gauge stations (more stations see Fig.~\ref{SI:SI_HRES_representative_station}); gray envelope denotes the spread across 100 ensemble members.}
\label{fig:HRES}
\end{figure}

PRIMER is not only effective for bias correction and downscaling of existing precipitation datasets, but also exhibits strong generalization. Figure~\ref{fig:HRES} illustrates PRIMER's ability to correct biases in previously unseen operational precipitation forecasts, using the ECMWF High-Resolution Forecast (HRES) as a representative example~\cite{buizza2018development}. Despite never being trained on HRES, PRIMER successfully corrects systematic biases in a typical precipitation event caused by typhoon landing (Fig.~\ref{fig:HRES}a,e). The ensemble mean of $P_{*}(x \mid \mathcal{O}_{\mathrm{HRES}})$ (Fig.~\ref{fig:HRES}b,f) aligns with HRES, while each member (Fig.~\ref{fig:HRES}c,g) captures a diverse range of physically plausible precipitation scenarios, reflecting the model's ability to encode meaningful uncertainty. Maps of $\Delta$CRPS (Fig.~\ref{fig:HRES}d,h) with widespread positive values (blue dots) indicate that PRIMER produces a reliable probabilistic ensemble system for HRES. 
These improvements arise from the Bayesian posterior sampling mechanism. By drawing samples from $P_{*}(x \mid \mathcal{O}_{\mathrm{HRES}})$, we effectively use the learned prior distribution $P_{*}(x)$—which has been calibrated to match gauge statistics—to adjust the original HRES forecasts. This process mitigates systematic biases inherent in the original HRES forecasts.
To illustrate these benefits more intuitively, we present time series at two representative gauge locations (Fig.~\ref{fig:HRES}i,j). The ensemble envelopes generated by PRIMER closely track observed precipitation peaks, offering a reliable uncertainty quantification for HRES. 
Taken together, these results underscore that PRIMER can perform physically consistent corrections on new forecast products without additional retraining (zero-shot adaptation) using its learned prior $P_{*}(x)$. This highlights the broader utility of PRIMER as a foundation model for downstream applications in precipitation prediction.

\section{Discussion}
\label{sec:Conclusion}


Existing precipitation datasets exhibit a persistent trade-off among spatial coverage, temporal resolution, and measurement accuracy, with no single data source simultaneously meeting these criteria. This fundamental limitation necessitates sophisticated fusion methods capable of integrating heterogeneous observations while overcoming the deficiencies from each source. Generative AI, particularly probabilistic diffusion models, offers a powerful approach by capturing intricate distribution of precipitation patterns. However, practical application has been severely limited by intrinsic challenges: the extreme variability and discontinuity of precipitation, and most important the paradox of establishing reliable priors from individually imperfect and incomplete datasets.

To overcome these barriers, we introduce PRIMER that directly represents precipitation as a continuous spatial field, seamlessly incorporating sparse gauge observations alongside dense gridded data without destructive interpolation. Our two-stage training procedure uniquely exploits the complementary strengths of different data sources: we initially establish robust climatological priors by leveraging broadly available gridded products, which, despite their wide coverage, exhibit considerable uncertainties. These priors are then refined using sparse but accurate gauge observations.
Benchmark evaluations highlight PRIMER’s capability to effectively integrate gauge observations with gridded data, providing localized realism without sacrificing large-scale spatial coherence—a significant innovation termed \textit{climatological jailbreak}. Experimental results demonstrate PRIMER’s superiority in bias correction and super-resolution enhancement of existing precipitation datasets, consistently outperforming priors derived solely from single-source observations. Furthermore, experiments reveal that incorporating additional gauge observations during posterior sampling process significantly enhances accuracy, underscoring PRIMER’s promising potential for future data assimilation applications. Crucially, PRIMER exhibits robust zero-shot generalization, maintaining physical consistency when applied to previously unseen operational forecasts. These findings underscore PRIMER’s substantial potential as an advanced, principled approach for reliable and physically coherent precipitation data fusion.

Despite the impressive performance of PRIMER, one notable limitation is the lack of high-quality, in-situ gauge observations over oceanic regions. Sparse instrument coverage in oceanic regions presents a challenge. Another limitation of our study is that we focused on precipitation fusion within China, rather than globally. This decision was primarily driven by the substantial computational demands of performing global precipitation fusion, which would require resources far beyond our current capacity, given that we only have access to two A100 40G GPUs. Additionally, precipitation is one of the most complex and discontinuous variables in the climate system, which provides a stringent benchmark for validating our methodology before considering its application to broader climate domains. 

Looking ahead, several compelling directions emerge from this study. First, integrating additional precipitation records~\cite{zhang2016multi, beck2017mswep}—such as more gauge observations or even advanced ground-based radar observations—could further improve the learned prior. Second, the PRIMER framework is intrinsically extensible. Its architecture naturally supports the integration of auxiliary meteorological variables—such as temperature, wind, and humidity—as additional channels. This opens a promising pathway toward holistic representations of the atmospheric state. In particular, applying this framework to future climate simulations from CMIP~\cite{stevens2024perspective, eyring2016overview} offers a unique opportunity. By training on Earth system model outputs across multiple scenarios, a generalized version could be developed to learn coherent, external-forcing-aware distributions of the Earth system state. Such a generative model would facilitate projections of future Earth system evolution and deepen our understanding of its underlying statistics. These directions highlight the broader potential behind PRIMER as a scalable and principled foundation for advancing Earth system science.

\clearpage
\newpage

\section{Method}
\label{sec:Method}

\subsection{Problem formulation}
\label{Formulation}

A general formulation of the precipitation data fusion task involves two key components: (1) constructing an informative prior distribution over the precipitation field, and (2) performing posterior inference given new observations. 

Let $x$ denote the target precipitation field. Different data sources—including gridded products such as satellite-derived and reanalysis datasets, as well as sparse in-situ gauge measurements—provide multiple versions of $x$, each with varying spatial coverage and accuracy. Our goal is to effectively leverage these heterogeneous sources to construct a unified prior distribution $P(x)$. This prior plays a central role, as it is expected to integrate statistical characteristics of each source through a balanced fusion.
A key innovation of this work lies in the design of a principled experimental framework for modelling such a prior.

Once an informative prior is established, posterior inference is conducted as new observational evidence $\mathcal{O}$ becomes available. Posterior distribution $P(x \mid \mathcal{O})$ can be factored into two components: the prior distribution $P(x)$, and the likelihood $P(\mathcal{O} \mid x)$.
Another innovation of our work is the effective implementation of posterior inference that balances the prior and the observations, ensuring the inferred precipitation field reflects both the climatological variability and the specific constraints provided by $\mathcal{O}$.
Consequently, this Bayesian framework naturally enables various downstream applications, such as super-resolution by conditioning on coarse-resolution data, bias correction by conditioning on biased estimates, and data assimilation by jointly conditioning on observations and background fields.

\subsection{Preliminary on diffusion models}
\label{Preliminary on diffusion models}
To construct a prior distribution, we employ score-based diffusion models within a principled learning strategy. To enable the model to distinguish between sources during training, we associate each sample with a corresponding entity embedding \(e_i\) (\(e_1 = (1, 0, 0), e_2 = (0, 1, 0), e_3 = (0, 0, 1)\))~\cite{guo2016entityembeddingscategoricalvariables}, which is injected into the model. This embedding functions as a source identifier, enabling the model to learn distinct priors for different data sources. Specifically, \(e_1\) corresponds to ERA5, \(e_2\) to IMERG, and \(e_3\) to gauge observations. 
Here, we first outline the foundations of the traditional diffusion framework before extending its conceptual scope.
The forward diffusion process evolves the data distribution into a tractable Gaussian through a stochastic differential equation (SDE)~\cite{song2020score, song2019generative, ho2020denoising, song2022denoisingdiffusionimplicitmodels}:
\begin{equation}
dx_t = f(x_t, t)\,dt + g(t)\,dW_t,
\end{equation}
where \( x_t \in \mathbb{R}^n \) is the state at time \(t\), \( f(x_t, t) \) is the drift function, and \(W_t\) is a standard Wiener process. To generate samples from priors, we solve the reverse-time SDE~\cite{song2020score, luo2022understandingdiffusionmodelsunified}:
\begin{equation}
dx_t = \left[ f(x_t, t) - g^2(t)\nabla_{x_t} \log P_\theta(x_t \mid e_i) \right] dt + g(t)\,dW_t,
\end{equation}
where the score function \(\nabla_{x_t} \log P_\theta(x_t \mid e_i)\) denotes the gradient of the log-density with respect to different sources. Since this score is intractable, we approximate it using a neural network $f_\theta$. 
We provide a theoretical justification for our proposed two-stage training strategy in SI Section~\ref{Theoretical justification for dual-source integration}.

\subsection{PRIMER}
\label{Infinite diffusion}
Traditional diffusion models typically rely heavily on U-Net architectures~\cite{ronneberger2015unetconvolutionalnetworksbiomedical}, which require inputs and outputs to be uniformly gridded data with fixed resolution. This architectural constraint limits their flexibility, particularly when processing discrete, sparse gauge observations. PRIMER utilizes a new framework inspired by recent theoretical advances~\cite{bond2023infty, pidstrigach2023infinitedimensionaldiffusionmodels, zhang2023functionaldiffusion, azizzadenesheli2024neural}, which generalizes diffusion models from finite-dimensional Euclidean space to an infinite-dimensional Hilbert space \( \mathcal{H} \), as illustrated in Figure~\ref{SI:Hilbert_space} (see SI Section~\ref{PRIMER_name_reason} for the origin of the name).
In this setting, each element \( x \in \mathcal{H} \) is a function \( x: \mathbb{R}^n \rightarrow \mathbb{R}^d \), where \( \mathbb{R}^n \) denotes coordinates and \( \mathbb{R}^d \) represents physical quantities. Both dense gridded data and sparse gauge observations are treated as partial realizations of an underlying function, allowing PRIMER to natively integrate heterogeneous records. Following Bond \emph{et al.}~\cite{bond2023infty}, we define \( \mathcal{H} \) as \( L^2([0,1]^n \rightarrow \mathbb{R}^d) \), where \( L^2 \) denotes the space of functions \( f \) such that \( \int_{[0,1]^n} |f(x)|^2 \, dx < \infty \).

\subsubsection{Mollification}
\label{subsubsection:Mollification}

While tempting, using white noise in the forward diffusion process poses a fundamental issue. Let \( \epsilon(\mathbf{c}) \) be a white noise where each \( \mathbf{c} \in \mathbb{R}^n \) is sampled independently from \( \mathcal{N}(0,1) \). For \( \epsilon \) to lie in the Hilbert space \( \mathcal{H} \), it must be square-integrable. However, \( \epsilon(\mathbf{c}) \) violates this, as its norm diverges. To address this, PRIMER applies a Gaussian kernel \( k \) to mollify the noise:
$
\xi(\mathbf{c}) = (k * \epsilon)(\mathbf{c}) = \int_{\mathbb{R}^n} k(\mathbf{c} - \mathbf{c}') \epsilon(\mathbf{c}') \, d\mathbf{c}'.
$
The resulting smoothed noise is square-integrable and thus belongs to \( \mathcal{H} \), as rigorously proven in SI~\ref{proof1}. Similarly, PRIMER also mollifies \( x_0 \), which ensures that $Lx_0$ inherit the same smoothness properties. In practice, this operation is implemented efficiently using Discrete Fourier Transformations (DFT). In Fourier space, mollification corresponds to:
$
\label{fourier_transformation}
\epsilon(\boldsymbol{\omega}) = e^{\|\boldsymbol{\omega}\|^2 t}\, \xi(\boldsymbol{\omega})
$,
where \( \boldsymbol{\omega} \in \mathbb{R}^n \) denotes the frequency vector, and \( t = \sigma^2 / 2 \), with \( \sigma \) being the standard deviation of kernel \( k \) (a detailed derivation is provided in SI~\ref{proof2}). Directly applying the inverse transformation is often numerically unstable, thus we employ Wiener filter, defined as \cite{bond2023infty, biemond2002iterative}:
$
\tilde{\epsilon}(\boldsymbol{\omega}) = \frac{e^{-\|\boldsymbol{\omega}\|^2 t}}{e^{-2\|\boldsymbol{\omega}\|^2 t} + \epsilon^2}\, \xi(\boldsymbol{\omega})
$,
where \( \epsilon \) is a small positive regularization parameter.

\subsubsection{Network architecture}
\label{Network architecture}
Neural Operators are capable for learning a map between two functional spaces~\cite{li2020neuraloperatorgraphkernel, kovachki2023neural, li2020fourier, azizzadenesheli2024neural}. Neural operators achieve discretization invariance by learning integral kernels parameterized via neural networks. Specifically, for an input function $x: \mathbb{R}^n \to \mathbb{R}^d$, with observations at $m$ distinct spatial locations, the operator $K(x; \theta)$ is defined as:
$$\bigl(K(x;\theta)x\bigr)(\mathbf{c}) = \int_{\mathbb{R}^n} \kappa_\theta\Bigl(\mathbf{c}, \mathbf{b}, x(\mathbf{c}), x(\mathbf{b})\Bigr) \, x(\mathbf{b})\, d\mathbf{b},$$
where $\kappa_\theta: \mathbb{R}^n \times \mathbb{R}^n \times \mathbb{R}^d \times \mathbb{R}^d \to \mathbb{R}$ is a kernel function parameterized by $\theta$, which captures complex non-local dependencies. However, applying Neural Operators like FNO~\cite{li2020fourier} directly to extensive spatial domains presents scalability and computational efficiency challenges~\cite{bond2023infty}. PRIMER implements a hybrid multi-scale architecture that synthesizes the strengths of Neural Operators and convolutional networks. 
PRIMER first processes the input feature \( x \in \mathbb{R}^{d \times m} \) using a series of SparseConvResBlocks, which primarily employ sparse depthwise convolutions~\cite{tang2022torchsparse}, producing updated features with shape \( \mathbb{R}^{D \times m} \), where \( D \gg d \). This embedding step projects low-dimensional input features into a higher-dimensional space, a crucial operation in deep learning that enables the model to capture richer representations. For the motivation behind SparseConvResBlock, see SI.~\ref{proof3}.
Since the features lie on an irregular set of discrete locations, we project them onto a coarse regular grid based on their spatial coordinates. This transformation aligns the features to a structured tensor.
A U-Net is applied to this grid to capture multi-scale context.
As we are ultimately interested in observations at the original irregular target locations, the processed grid features are reprojected to these coordinates via bilinear interpolation, yielding a feature tensor of shape \( \mathbb{R}^{D \times m} \).
Finally, a subsequent series of SparseConvResBlocks refines the features to produce the output tensor of shape \( \mathbb{R}^{d \times m} \).
For details about network architecture, see SI~\ref{SI_Network_architecture}

%
%
%
%

\subsubsection{Model training}

The model is optimized by minimizing a simplified denoising objective~\cite{ho2020denoising, song2020score, bond2023infty} (derivation provided in SI Section~\ref{Training and inference algorithm pseudocode}):
\begin{equation}
\mathcal{L} = \mathbb{E}_{t} \left[\|f_\theta(x_t, t, e_i) - \xi\|_{\mathcal{H}}^2\right],
\end{equation}
where $x_t$ denotes the noisy input at time step $t$, $e_i$ represents the embedding of data source, $\xi$ is the ground-truth noise, and $\|\cdot\|_{\mathcal{H}}$ denotes the loss norm defined in Hilbert space $\mathcal{H}$.
We adopt a two-stage training procedure. In Stage~1, the model is jointly trained on ERA5 ($e_1$) and IMERG ($e_2$) data. 
In Stage~2, we specialize the pretrained model to sparse gauge observations ($e_3$), following a personalization-inspired strategy akin to DreamBooth~\cite{ruiz2023dreambooth}. 
Specifically, we fine-tune the model using a shared-weight strategy, where training samples are proportionally drawn from multiple data sources. The total loss is computed as:
\begin{equation}
\mathcal{L}_{\text{fine-tuning}} = \alpha_1 \mathcal{L}_{\text{ERA5}} + \alpha_2 \mathcal{L}_{\text{IMERG}} + \alpha_3 \mathcal{L}_{\text{gauge}},
\end{equation}
with weights $\alpha_1 = 0.1$, $\alpha_2 = 0.4$, and $\alpha_3 = 0.5$. This strategy enables the model to preserve global climatological priors while adapting to high-fidelity signals, effectively grounding the generative manifold towards real-world observations.

All models are optimized using the AdamW optimizer with $\beta_1 = 0.9$, $\beta_2 = 0.99$, and weight decay of $4 \times 10^{-6}$.
The full training and inference pipelines are summarized in SI Algorithm~\ref{alg:primer_training} and SI Algorithm~\ref{alg:inference}, with an overview schematic shown in SI Fig.~\ref{SI:algorithm_illustration}. 
For the configuration of the hyperparameters, see SI Section~\ref{Choice of experiment parameter}.

\subsection{Posterior sampling}
\label{Posterior sampling}

In tasks such as bias correction, downscaling, and data assimilation, the objective is to infer an unknown target state $x$ from observations $\mathcal{O}$. A Bayesian framework enables the incorporation of prior knowledge through a prior distribution $P(x)$, facilitating posterior inference via Bayes' theorem: $P(x|\mathcal{O}) \propto P(\mathcal{O}|x)P(x).$ When employing PRIMER as priors, the standard reverse-time SDE can be adapted to sample from the posterior distribution. The modified reverse diffusion process takes the form:
\begin{equation}
    dx_t = \left[ f(x_t, t) - g^2(t) \left(\nabla_{x_t} \log P_\theta(x_t \mid e_i) + \nabla_{x_t} \log P_\theta(\mathcal{O} \mid e_i, x_t)\right)\right] dt + g(t)\,dW_t.
\end{equation}

This formulation requires two key components: the time-dependent score function $\nabla_{x_t} \log P_\theta(x_t \mid e_i)$, which can be approximated by a trained score network; and the gradient of the likelihood $\nabla_{x_t} \log P_\theta(\mathcal{O} \mid e_i, x_t)$, which remains challenging to estimate due to the generally intractable dependency between $\mathcal{O}$ and $x_t$. Several recent studies have proposed various strategies to address posterior sampling within the diffusion framework~\cite{zheng2025inversebenchbenchmarkingplugandplaydiffusion, chung2024diffusionposteriorsamplinggeneral, daras2024surveydiffusionmodelsinverse}. In light of the characteristics of our problem setting, we adopt two representative approaches: Inpainting~\cite{chao2024learning, lugmayr2022repaint, zhang2023towards} and SDEdit~\cite{meng2022sdeditguidedimagesynthesis}. 

Inpainting in diffusion models reconstructs unobserved regions by conditioning on partial observations \(\mathcal{O}\). A binary mask \(\mathbf{m}\) indicates observed entries (\(m_i = 1\) if observed). At each reverse-time step \(t\), a denoised estimate \(\hat{x}_t\) is first computed. To enforce consistency with known observations, we blend the latent state using
\[
x_t = \mathbf{m} \odot q(x_t | \mathcal{O}) + (1 - \mathbf{m}) \odot \hat{x}_t,
\]
where \(\odot\) denotes element-wise multiplication. The term \(q(x_t | \mathcal{O})\) is constructed by applying the same forward noise process to \(\mathcal{O}\); that is, for each observed entry, we simulate its noisy counterpart at step \(t\) under the forward SDE. This blending operation preserves observed values while allowing the model to impute missing regions, approximating the posterior distribution \(p(x | \mathcal{O})\).
SDEdit can be viewed as a special case of inpainting where the entire input field is treated as observed, i.e., \(\mathbf{m} = \mathbf{1}\). However, a key distinction lies in its use of a noise level parameter \(\tau\), which determines the strength of forward noise applied to the input before denoising. This parameter controls the extent to which the model is allowed to deviate from the original input, balancing fidelity and diversity. To select an appropriate \(\tau\), we conduct a sensitivity analysis on IMERG precipitation data for 13 June 2016 at 23:00 UTC.
For each noise level from 0.1 to 0.9 in steps of 0.1, we generate an ensemble of 50 samples from posterior $P_{*}(x \mid \mathcal{O}_{\mathrm{IMERG}})$ and compute both the ensemble mean root mean square error (RMSE) and the continuous ranked probability score (CRPS) over 50 repeated subsampling trials, each selecting 10 members randomly. As shown in SI Fig.~\ref{fig:SI_determine_alpha}, performance improves with increasing \(\tau\) up to around 0.6, beyond which both RMSE and CRPS begin to deteriorate. 
This suggests an optimal trade-off at 0.6 noise levels, where PRIMER maintains sufficient variability to explore plausible outcomes while preserving alignment with observational constraints.

\subsection{Statistical methods}

In the main text, we analyze the statistical properties of different prior distributions, including $P_{\mathrm{ERA5}}(x)$, $P_{\mathrm{IMERG}}(x)$, and the updated prior $P_{*}(x)$. These priors are then used for posterior sampling. To isolate the effect of the prior, all posterior distributions are conditioned on the same observational evidence $\mathcal{O}$. As a result, differences in posterior accuracy primarily reflect differences in the quality of the corresponding priors.
To evaluate the performance of each posterior distribution, we adopt the following evaluation method, described in detail below.

\subsubsection{Evaluation metrics}

\paragraph{Deterministic accuracy.}  
To assess the accuracy of the ensemble mean forecast, we report the Mean Absolute Error (MAE) and the Pearson Correlation Coefficient (PCC). MAE captures the average absolute deviation between the predicted ensemble mean \(\hat{x}\) and the observed value \(x\):
\begin{equation}
\mathrm{MAE} = \frac{1}{N} \sum_{i} \left| \hat{x}_i - x_i \right|.
\end{equation}
where $i$ indexes the grid points corresponding to the gauge locations. PCC measures the linear association between predicted and observed spatial fields:
\begin{equation}
\mathrm{PCC} = \frac{ \sum_i (\hat{x}_i - \bar{\hat{x}})(x_i - \bar{x}) }{ \sqrt{ \sum_i (\hat{x}_i - \bar{\hat{x}})^2 } \sqrt{ \sum_i (x_i - \bar{x})^2 } }.
\end{equation}
Here, \( \bar{\hat{x}} \) and \( \bar{x} \) denote the spatial means of the predicted and observed fields, respectively. High PCC indicates strong spatial agreement.

\paragraph{Probabilistic skill.} 
\label{CRPS} 
We use the Continuous Ranked Probability Score (CRPS)~\cite{gneiting2007strictly}, a proper scoring rule that measures the quality of probabilistic forecasts by comparing the predicted cumulative distribution function (CDF) \(F\) with the observation \(y\). It is defined as:
\begin{equation}
\mathrm{CRPS}(F, y) = \int_{-\infty}^{\infty} \left(F(x) - \mathbf{1}_{\{x \geq y\}}\right)^2 \, dx,
\end{equation}
where \(\mathbf{1}_{\{x \geq y\}}\) is the Heaviside step function centered at \(y\). Lower CRPS value indicates a better-calibrated ensemble system.

\subsubsection{Evaluation tool}
\paragraph{Spatial lagged correlation coefficient}
\label{spatial_autocorrelation}

We evaluate the spatial dependency of a geophysical field $x \in \mathbb{R}^{H \times W}$ by computing its correlation with spatially shifted copies.
For each fixed offset $(\Delta i, \Delta j)$, we compute the Pearson correlation between $x$ and its lagged version $x_{\Delta i, \Delta j}$, using only the overlapping valid gauge observations.
This metric quantifies the degree to which values at one location are linearly correlated with values at a fixed spatial offset (lag) from that location, thus capturing the spatial dependency structure.

%

\paragraph{Empirical Orthogonal Function (EOF) decomposition}
\label{EOF_decomposition}

Given an anomaly matrix $x \in \mathbb{R}^{N \times T}$, where each row corresponds to spatial points and each column represents time instances, EOF decomposition factorizes $x$ via~\cite{hannachi2004primer}:
\begin{equation}
x = L Y,
\end{equation}
where $L \in \mathbb{R}^{N \times N}$ contains orthonormal spatial modes (EOFs), and $Y \in \mathbb{R}^{N \times T}$ holds the corresponding time coefficients (principal components). EOFs are derived as eigenvectors of the covariance matrix $S = \frac{1}{N-1} x x^\top$, arranged in decreasing order of eigenvalues, which represent the explained variance of each mode.

\paragraph{Radially averaged power spectral density (RAPSD)}
\label{Radially Averaged Power Spectral Density (RAPSD)}

To quantify spatial variability, we compute the radially averaged power spectral density (RAPSD) using the open-source Pysteps library~\cite{pulkkinen2019pysteps}. Given a 2D scalar field $f(x, y) \in \mathbb{R}^{H \times W}$, its discrete Fourier transform is
$
F(k_x, k_y) = \sum_{x=0}^{H-1} \sum_{y=0}^{W-1} f(x, y) \, e^{-2\pi i \left( \frac{k_x x}{H} + \frac{k_y y}{W} \right)},
$
and the corresponding power spectral density is

\begin{equation}
P(k_x, k_y) = \frac{1}{HW} \left| F(k_x, k_y) \right|^2.
\end{equation}

RAPSD is obtained by averaging $P(k_x, k_y)$ over annular bins of constant radial wavenumber $k = \sqrt{k_x^2 + k_y^2}$:

\begin{equation}
\mathrm{RAPSD}(k) = \frac{1}{N_k} \sum_{(k_x, k_y) \in \mathcal{A}_k} P(k_x, k_y),
\end{equation}

where $\mathcal{A}_k$ denotes the components in each bin. We express RAPSD as a function of wavelength $\lambda = 1/k$ to highlight scale-dependent variability.

\subsection{Data}
\label{data_description}
Pretraining uses two gridded datasets: Integrated Multi-satellitE Retrievals for GPM (IMERG)~\cite{huffman2015nasa} and ERA5~\cite{hersbach2020era5}. IMERG provides global precipitation estimates at \SI{0.1}{\degree} spatial and 30-minute temporal resolutions, derived from GPM satellite observations. 
To match ERA5's hourly resolution, pairs of consecutive 30-minute IMERG intervals are averaged to produce hourly estimates.
The study focuses on East Asia (20-45°N, 100-125°E), a region of high population density. After cropping, IMERG data form $250 \times 250$ grids, with 2000-2020 (excluding 2016) used for training. ERA5, from ECMWF, provides hourly precipitation at \SI{0.25}{\degree} resolution, yielding $100 \times 100$ grids over the same domain. Both datasets are log-transformed as \( x' = \log_{10}(0.1 + x) \) and standardized using IMERG statistics. 
For fine-tuning, we use dataset from Shen et al.~\cite{shen2014high}, constructed using over 30{,}000 Automatic Weather Stations (AWS) across China. The gridded dataset has a spatial resolution of 0.1$^\circ$ and a temporal resolution of 1 hour, covering 2015 and 2017 for training, and 2016 for testing. We select grid cells with at least one assimilated AWS observation for training, and use a subset with no fewer than four AWS observations as ground truth for evaluation, assuming higher reliability (see SI Fig.~\ref{fig:SI_gauges_info} for the spatial distribution of these gauges).
After identical cropping and preprocessing, the data are organized as two arrays: \texttt{gauge\_observation} $(N, 1)$ for precipitation intensity and \texttt{gauge\_coordinate} $(N, 2)$ for location (longitude, latitude), both of which are input into the model during fine-tuning.

IFS HRES is ECMWF’s ﬂagship deterministic highresolution model and is widely regarded as one of the best physics-based numerical-weather-forecast models in the world~\cite{rasp2020weatherbench, olivetti2024data}.
HRES produces hourly forecasts at a 0.1° horizontal resolution. It is included in our experiments to demonstrate PRIMER’s strong generalization capability even on datasets it was not trained on. 
For consistency, HRES forecasts undergo the same cropping and preprocessing steps as IMERG.

\clearpage
\newpage

\backmatter

\section*{Declarations}

\begin{itemize}

\item Data availability: ERA5 reanalysis were obtained from the Copernicus Climate Change Service's Climate Data Store (CDS) (\url{https://cds.climate.copernicus.eu}). For the quickest access, the WeatherBench2 data archive provides an efficient alternative (\url{https://console.cloud.google.com/storage/browser/weatherbench2}). The IMERG data can be accessed from \url{https://disc.gsfc.nasa.gov/datasets/GPM_3IMERGHH_07/summary?keywords=imerg}. Gauge observations were provided from the China Meteorological Administration (CMA) under license. However, restrictions apply to the availability of these data, which were used under license for the present study. Data are available from the authors upon reasonable request and with permission from the CMA. A small subset of gauge observations will be made available on GitHub to facilitate reproducibility and support code debugging. The high-resolution forecast data (HRES) from the Integrated Forecasting System (IFS) used in this study are produced by the European Centre for Medium-Range Weather Forecasts (ECMWF). For more detailed information on HRES access, please refer \url{https://www.ecmwf.int/en/forecasts/datasets/set-i}.

\item Code availability: The code implementing PRIMER will be available on the GitHub repository. Model configurations and training scripts used in this study will be made publicly available upon acceptance of this work.
\item Acknowledgemenrs: This work was supported by the National Natural Science Foundation of China (42130603).
 
\end{itemize}


\newpage

\begin{appendices}

\section{Why “PRIMER”}
\label{PRIMER_name_reason}
The name PRIMER (Precipitation Records Infinite MERging) is deliberately chosen—not only as an acronym, but also as a metaphor. In English, a ``primer" refers to a preparatory coating applied before the final layer of paint or makeup, ensuring better adhesion, durability, and refinement. Similarly, our framework first performs extensive pretraining on gridded products like ERA5 and IMERG before fine-tuning with sparse, high-quality gauge observations. This staged approach allows PRIMER to seamlessly merge multi-sources of records. We envision this method as a general-purpose ``foundation layer'' for geoscientific modeling—particularly valuable in domains where accurate downstream tasks rely on the fusing of heterogeneous records.

\section{Method details}
\label{Method details}

\subsection{Theoretical justification for dual-source integration}
\label{Theoretical justification for dual-source integration}

We provide a theoretical justification for our proposed two-stage integration framework by drawing parallels with recent advances in diffusion theory~\cite{daras2024noisyimageworthdata, daras2023ambient, daras2023consistent, daras2024consistent, dieleman2024spectral}. Specifically, we aim to establish an upper bound on the Wasserstein-1 distance~\cite{panaretos2019statistical} between the true precipitation distribution $\mathcal{P}_{\mathrm{true}}$ and the learned model distribution $\hat{\mathcal{P}}$. The model is trained in two stages: first on a large, noisy gridded dataset $\mathcal{D}$, and subsequently fine-tuned on a sparse but high-fidelity gauge dataset $\mathcal{D}^*$. In this setup:
\begin{itemize}
    \item $\mathcal{D}$ comprises low-quality, relatively high-uncertainty samples;
    \item $\mathcal{D}^*$ comprises accurate, relatively low-uncertainty gauge observations.
\end{itemize}

Assume we observe precipitation samples $X_1, \dots, X_n$, independently drawn from the following generative process:
\begin{equation}
X_i \sim \mathcal{P}_{\mathrm{true}} * \mathcal{N}(0, \sigma_i^2 I),
\end{equation}
where $\mathcal{P}_{\mathrm{true}} = \sum_{j=1}^k w_j \delta_{\mu_j}$ is a finite $k$-component mixture of point masses (or equivalently, a degenerate Gaussian mixture). Each $\delta_{\mu_j}$ denotes a representative precipitation mode with weight $w_j$, and $\sigma_i$ captures the noise level of the $i$-th observation. While this discrete formulation does not fully capture the full distribution of precipitation fields, it provides a tractable approximation that enables analytical insight into complex precipitation distributions, in line with common practice in diffusion-based modeling. Our objective is to learn a distribution $\hat{\mathcal{P}}$ that closely approximates $\mathcal{P}_{\mathrm{true}}$ by minimizing their Wasserstein-1 distance:
\begin{equation}
W_1(\mathcal{P}_{\mathrm{true}}, \hat{\mathcal{P}})
= \min_{C(\mathcal{P}_{\mathrm{true}}, \hat{\mathcal{P}})}
\mathbb{E}_{(X, X') \sim C} [\| X - X' \|],
\end{equation}
where $C$ is a valid coupling between $\mathcal{P}_{\mathrm{true}}$ and $\hat{\mathcal{P}}$ with marginals equal to $\mathcal{P}$ and $\hat{\mathcal{P}}$.

By invoking the theoretical framework established in~\cite{daras2024noisyimageworthdata}, particularly Theorem 4.2, we obtain the following upper bound. Let $n$ be the total number of samples (from both $\mathcal{D}$ and $\mathcal{D}^*$), $d$ the dimensionality of the data space, and $k$ the number of mixture components. Then, under mild regularity assumptions, there exists a procedure returning $\hat{\mathcal{P}}$ such that with probability at least $1 - \delta$:
\begin{equation}
\label{eq:theory_bound}
\begin{aligned}
W_1(\mathcal{P}_{\mathrm{true}}, \hat{\mathcal{P}})
\leq C \Bigg(
& k \left(
\frac{d + \log(1/\delta)}
{\sum_{i=1}^n 1/\sigma_i^4}
\right)^{1/4}
+
k^3 \left(
\frac{k \log k + \log(1/\delta)}
{\sum_{i=1}^n 1/\sigma_i^{4k - 2}}
\right)^{1/(4k - 2)}
\Bigg),
\end{aligned}
\end{equation}
The estimation error bound naturally decomposes into two principal components. The first term reflects the \emph{dimensionality reduction error}, arising from the challenge of projecting high-dimensional precipitation fields (with dimension $d$) into a lower-dimensional subspace of $k$ modes. The second term quantifies the \emph{low-dimensional estimation error}, which captures the precision of parameter estimation within this $k$-dimensional space. This decomposition mirrors the two-stage process used in our framework and aligns closely with recent analyses in ambient diffusion~\cite{daras2024noisyimageworthdata}. In the first stage, the model compresses the data into a reduced representation, where the estimation accuracy depends on the effective sample size, approximately represented by $\sum_{i=1}^n 1/\sigma_i^4$. Here, the gridded dataset $\mathcal{D}$—although characterized by high noise levels $\sigma_i$—offers substantial benefit due to its extensive spatial  coverage and large number of samples. While noise rapidly degrades high-frequency information, it affects low-frequency components (i.e., structural patterns) to a lesser degree. Consequently, gridded datasets remain valuable for capturing the overall structure of the precipitation field, making $\mathcal{D}$ instrumental in reducing the dimensionality of the problem. In the second stage, the model performs fine-grained estimation within the reduced space, where performance becomes sensitive to uncertainties. The corresponding term in the bound depends on $\sum_{i=1}^n 1/\sigma_i^{4k - 2}$, emphasizing the critical role of low-uncertainty gauge observations. Although our gauge dataset $\mathcal{D}^*$ is sparse, its markedly smaller noise variances make it disproportionately influential in this stage. Analogous to the clean data in~\cite{daras2024noisyimageworthdata}, $\mathcal{D}^*$ effectively preserves high-frequency components. This theoretical framing highlights a key insight: low-quality (noisy) data are primarily useful for capturing structural information and aiding dimensionality reduction, while high-quality (clean) data are essential for refining local accuracy. By strategically combining $\mathcal{D}$ and $\mathcal{D}^*$, our framework balances this trade-off, leveraging the complementary strengths of each data source to robustly approximate the underlying precipitation distribution. To illustrate this point concretely, consider a simplified case where $p$ fraction of the dataset is $D^*$ (gauge observations), and $(1 - p)$ fraction is $D$  (reanalysis or satellite retrievals). Then, Eq.~\eqref{eq:theory_bound} simplifies to (see corollary 4.3 in~\cite{daras2024noisyimageworthdata}):
\begin{equation}
\begin{aligned}
W_1(\mathcal{P}_{\mathrm{true}}, \hat{\mathcal{P}})
\leq C \Bigg(
& k \left(
\frac{d + \log(1/\delta)}
{n \left( p + (1 - p)/\sigma^4 \right)}
\right)^{1/4}
+
k^3 \left(
\frac{k \log k + \log(1/\delta)}
{n \left( p + (1 - p)/\sigma^{4k - 2} \right)}
\right)^{1/(4k - 2)}
\Bigg),
\end{aligned}
\end{equation}
revealing that the high-uncertainty samples are down-weighted by $1/\sigma^4$ (dimensionality reduction) and, more substantially, by $1/\sigma^{4k - 2}$ (fine-grained estimation).

Overall, this theoretical framework underpins the rationale for our two-stage strategy:
\begin{enumerate}
    \item Pretraining on $\mathcal{D}$ exploits its large sample size to establish robust large-scale spatial structure (reflected in the first term of the bound);
    \item Fine-tuning on $\mathcal{D}^*$ leverages its high-fidelity observations to minimize local estimation error (reflected in the second term).
\end{enumerate}

\subsection{Proof that mollified white noise belongs to \( L^2 \) space}
\label{proof1}

As established in the main text, white noise \( \epsilon(\mathbf{c}) \) is not an element of space \( \mathcal{H} = L^2([0,1]^n \rightarrow \mathbb{R}^d) \). In this section, we formally show that its mollified version, obtained via convolution with a Gaussian kernel, is square-integrable and hence admissible within the Hilbert space.

\begin{proof}

Let \( G(\mathbf{c}) = \frac{1}{(2\pi\sigma^2)^{n/2}} \, e^{-\frac{\|\mathbf{c}\|^2}{2\sigma^2}} \) denote a Gaussian kernel with variance \( \sigma^2 \). We define the mollified signal as the convolution:
\[
\xi(\mathbf{c}) = (\epsilon * G)(\mathbf{c}).
\]

By the convolution theorem, the Fourier transform of \( \xi \) is the product of the transforms of its components:
\[
\mathcal{F}[\xi](\boldsymbol{\omega}) = \mathcal{F}[\epsilon](\boldsymbol{\omega}) \cdot \mathcal{F}[G](\boldsymbol{\omega}),
\]
where \( \boldsymbol{\omega} \in \mathbb{R}^n \) denotes the frequency vector. The Fourier transform of the Gaussian is again Gaussian:
\[
\mathcal{F}[G](\boldsymbol{\omega}) = e^{-\sigma^2 \|\boldsymbol{\omega}\|^2 / 2}.
\]

Applying Parseval's theorem, the squared \( L^2 \) norm of \( \xi \) in physical space is equal to that in frequency space:
\[
\|\xi\|^2 = \int_{\mathbb{R}^n} |\xi(\mathbf{c})|^2 \, d\mathbf{c} = \int_{\mathbb{R}^n} |\mathcal{F}[\xi](\boldsymbol{\omega})|^2 \, d\boldsymbol{\omega}.
\]

Substituting the frequency-domain representation:
\[
\|\xi\|^2 = \int_{\mathbb{R}^n} |\mathcal{F}[\epsilon](\boldsymbol{\omega})|^2 \cdot e^{-\sigma^2 \|\boldsymbol{\omega}\|^2} \, d\boldsymbol{\omega}.
\]

Assuming \( \epsilon(\mathbf{c}) \) is white noise, its power spectral density is constant in expectation: \( |\mathcal{F}[\epsilon](\boldsymbol{\omega})|^2 = C \). Thus:
\[
\|\xi\|^2 \propto \int_{\mathbb{R}^n} e^{-\sigma^2 \|\boldsymbol{\omega}\|^2} \, d\boldsymbol{\omega}.
\]

This is a standard Gaussian integral over \( \mathbb{R}^n \), yielding:
\[
\int_{\mathbb{R}^n} e^{-\sigma^2 \|\boldsymbol{\omega}\|^2} \, d\boldsymbol{\omega} = \left( \frac{\pi}{\sigma^2} \right)^{n/2} < \infty.
\]

Therefore, the mollified signal \( \xi(\mathbf{c}) \) has finite energy and lies in \( L^2([0,1]^n \rightarrow \mathbb{R}^d) \), satisfying the requirement for inclusion in the Hilbert space \( \mathcal{H} \) used in PRIMER.

\end{proof}

\subsection{Proof of Fourier-domain relationship between original and mollified fields}
\label{proof2}

We aim to show that mollifying a signal \( x(\mathbf{c}) \in \mathcal{H} \) by convolving it with a Gaussian kernel results in a Fourier-domain relation:
\[
\hat{x}(\boldsymbol{\omega}) = e^{\|\boldsymbol{\omega}\|^2 t} \hat{h}(\boldsymbol{\omega}),
\]
where \( \hat{x}(\boldsymbol{\omega}) \) and \( \hat{h}(\boldsymbol{\omega}) \) denote the Fourier transforms of the original and mollified fields, respectively, and \( t = \sigma^2 / 2 \) is determined by the kernel \( \sigma \). This relation is central to the spectral manipulation used in PRIMER.

\begin{proof}
Consider the Gaussian kernel in \( \mathbb{R}^n \):
\[
k(\mathbf{c}) = \frac{1}{(2\pi \sigma^2)^{n/2}} \, e^{-\frac{\|\mathbf{c}\|^2}{2\sigma^2}}, \quad \mathbf{c} \in \mathbb{R}^n.
\]

Its Fourier transform is given by:
\[
\hat{k}(\boldsymbol{\omega}) = \int_{\mathbb{R}^n} k(\mathbf{c}) \, e^{-i \boldsymbol{\omega} \cdot \mathbf{c}} \, d\mathbf{c}.
\]

Substituting for \( k(\mathbf{c}) \):
\[
\hat{k}(\boldsymbol{\omega}) = \frac{1}{(2\pi \sigma^2)^{n/2}} \int_{\mathbb{R}^n} e^{-\frac{\|\mathbf{c}\|^2}{2\sigma^2}} e^{-i \boldsymbol{\omega} \cdot \mathbf{c}} \, d\mathbf{c}.
\]

Complete the square in the exponent:
\[
-\frac{\|\mathbf{c}\|^2}{2\sigma^2} - i \boldsymbol{\omega} \cdot \mathbf{c}
= -\frac{1}{2\sigma^2} \left( \|\mathbf{c} + i\sigma^2 \boldsymbol{\omega}\|^2 + \sigma^4 \|\boldsymbol{\omega}\|^2 \right).
\]

Thus:
\[
\hat{k}(\boldsymbol{\omega}) = \frac{1}{(2\pi \sigma^2)^{n/2}} \, e^{-\frac{\sigma^2 \|\boldsymbol{\omega}\|^2}{2}} \int_{\mathbb{R}^n} e^{-\frac{\|\mathbf{c} + i\sigma^2 \boldsymbol{\omega}\|^2}{2\sigma^2}} d\mathbf{c}.
\]

The integral evaluates to the normalization constant due to translation invariance:
\[
\int_{\mathbb{R}^n} e^{-\frac{\|\mathbf{c} + i\sigma^2 \boldsymbol{\omega}\|^2}{2\sigma^2}} d\mathbf{c} = (2\pi \sigma^2)^{n/2}.
\]

Therefore:
\[
\hat{k}(\boldsymbol{\omega}) = e^{-\frac{\sigma^2 \|\boldsymbol{\omega}\|^2}{2}}.
\]

Setting \( \sigma = \sqrt{2t} \) gives:
\[
\hat{k}(\boldsymbol{\omega}) = e^{-\|\boldsymbol{\omega}\|^2 t}.
\]

For the mollified signal \( h(\mathbf{c}) = (x * k)(\mathbf{c}) \), the convolution theorem implies:
\[
\hat{h}(\boldsymbol{\omega}) = \hat{x}(\boldsymbol{\omega}) \cdot \hat{k}(\boldsymbol{\omega}) = \hat{x}(\boldsymbol{\omega}) \cdot e^{-\|\boldsymbol{\omega}\|^2 t},
\]

so rearranging yields:
\[
\hat{x}(\boldsymbol{\omega}) = e^{\|\boldsymbol{\omega}\|^2 t} \cdot \hat{h}(\boldsymbol{\omega}).
\]

\end{proof}

\begin{figure}[htbp]
    \centering
 \includegraphics[width=\textwidth,page=1]{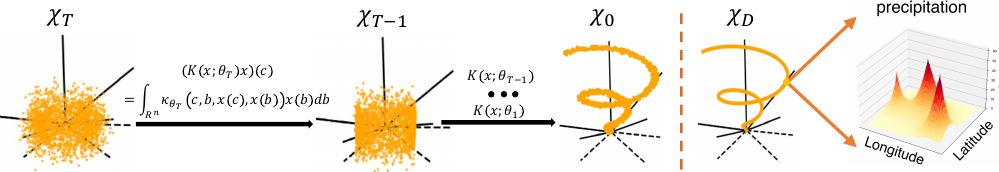} 
    \caption{\textbf{Conceptual illustration of the denosing process in the PRIMER.} Each point within the spaces $\chi_T$, $\chi_{T-1}$, $\dots$, $\chi_0$, and $\chi_D$ represents a function residing in an infinite-dimensional Hilbert space, as indicated by the axes. Through iterative transformations governed by neural operators $K(x;\theta_t)$ parameterised by kernel $\kappa_{\theta_t}$, PRIMER progressively transforms the initial $\chi_T$ toward the targeted distribution $\chi_0$, closely approximating the desired distribution represented by $\chi_D$. The rightmost panel visually illustrates such a function. The proximity between the distributions of $\chi_0$ and $\chi_D$ highlights PRIMER's capability in modeling the phase space (distribution), serving as an useful prior.}

    \label{SI:Hilbert_space}
\end{figure}

\begin{figure}[htbp]
    \centering
    \includegraphics[width=\textwidth,page=1]{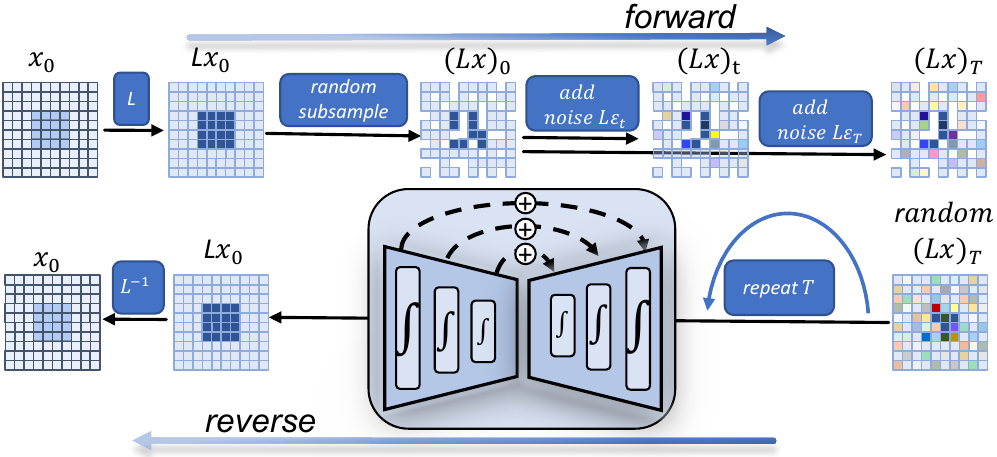} 
    \caption{\textbf{Schematic illustration of the algorithm.} The upper row depicts the forward process: starting with an initial state $x_0$, a smoothing gaussian kernel $L$ is applied, followed by random subsampling to create $(Lx)_0$. Progressive noise addition generates intermediate states $(Lx)_t$ and the final state $(Lx)_T$. The lower row shows the reverse process: beginning with the noisy observation $(Lx)_T$, the neural network iteratively denoises the signal through $T$ repetitions, ultimately recovering $Lx_0$. The inverse operator $L^{-1}$ then reconstructs the original signal $x_0$ using Wiener filter.}

    \label{SI:algorithm_illustration}
\end{figure}

\subsection{Training and inference algorithm pseudocode}
\label{Training and inference algorithm pseudocode}

We introduce the transition distribution~\cite{bond2023infty} in the forward process:
\begin{equation}
q(x_t \mid x_0) = \mathcal{N}\left(x_t; \sqrt{\bar{\alpha}_t} \,L x_0, (1 - \bar{\alpha}_t) L L^*\right),
\end{equation}
where \( L \) denotes the mollification operator and  constant coefficient \( \bar{\alpha}_t \in [0,1] \) controls the balance between signal preservation and noise injection. The corresponding posterior distribution can be derived analytically~\cite{song2020score, bond2023infty}:
\begin{equation}
q(x_{t-1} \mid x_t, x_0) = \mathcal{N}\left(x_{t-1}; \tilde{\mu}_t(x_t, x_0), \tilde{\beta}_t L L^*\right),
\end{equation}
where the mean and variance are given by
\begin{align}
\tilde{\mu}_t(x_t, x_0) 
&= \frac{1}{\sqrt{\alpha_{t}}} \left( x_{t} - \frac{1 - \alpha_{t}}{\sqrt{1 - \bar{\alpha}_{t}}} \xi \right), \quad \xi \sim \mathcal{N}(0, L L^*), \\
\tilde{\beta}_t 
&= \frac{1 - \bar{\alpha}_{t-1}}{1 - \bar{\alpha}_t} \beta_t.
\end{align}

We use a neural network \( f_\theta \) to predict $\xi$. The reverse transition is defined as~\cite{song2020score, bond2023infty}:
\begin{equation}
p_\theta(x_{t-1} \mid x_t) = \mathcal{N}\left(x_{t-1}; \mu_\theta(x_t, t), \tilde{\beta}_t L L^*\right),
\label{p_theta}
\end{equation}
with the predicted mean given by
\begin{equation}
\mu_{\theta}(x_t, t) = \frac{1}{\sqrt{\alpha_{t}}} \left[x_{t} - \frac{1 - \alpha_{t}}{\sqrt{1 - \bar{\alpha}_{t}}} f_{\theta}(x_t, t)\right].
\end{equation}

Thus, PRIMER is trained by minimizing the following simplified objective~\cite{ho2020denoising, song2020score, bond2023infty}:
\begin{equation}
\mathcal{L}_{\text{simple}} =  \mathbb{E}_{t} \left[\|\mu_{\theta}(x_t, t) - \tilde{\mu}_t(x_t, x_0) \|_{\mathcal{H}}^2\right] = \mathbb{E}_{t} \left[\|f_\theta(x_t, t) - \xi\|_{\mathcal{H}}^2\right].
\end{equation}

\begin{algorithm}[htbp]
\caption{Training Procedure of PRIMER}
\label{alg:primer_training}
\begin{algorithmic}[1]
\Require Gauge observations \( x_0 \in \mathbb{R}^{M \times d} \) (\( M \) gauges, each with a state vector in \( \mathbb{R}^d \)), gauge coordinates \( C \in \mathbb{R}^{M \times n} \) (spatial locations in \( \mathbb{R}^n \)), Gaussian mollifier kernel \( k \), diffusion schedule \( \bar{\alpha}_t \)

\State Sample white noise \( \varepsilon : \mathbb{R}^n \to \mathbb{R}^d \sim \mathcal{N}(0, I) \)

\ForAll{coordinates \( c \in C \)}
    \State Compute mollified noise: \( \xi(c) \gets (L\varepsilon)(c) = \int_{\mathbb{R}^n} k(c - c') \varepsilon(c') \, dc' \)
    \State Compute mollified data: \( (Lx_0)(c) = \int_{\mathbb{R}^n} k(c - c') x_0(c') \, dc' \)
    \State Compute diffused sample at time \( t \): 
    \(
    x_t(c) \gets \sqrt{\bar{\alpha}_t} \cdot (Lx_0)(c) + \sqrt{1 - \bar{\alpha}_t} \cdot \xi(c)
    \)
\EndFor

\State Predict mollified noise using neural network: \( \hat{\xi} \gets f_{\theta}(x_t, t) \)

\State Compute loss: \( \mathcal{L} \gets \| \hat{\xi} - \xi \|_{C}^2 \)

\State Update model parameters \( \theta \) using gradient descent on \( \mathcal{L} \)

\end{algorithmic}
\end{algorithm}

\begin{algorithm}[htbp]
\caption{Inference Procedure of PRIMER}
\label{alg:inference}
\begin{algorithmic}[1]
\Require Coordinates \( C \in \mathbb{R}^{M \times n} \)  (\( M \) gauges, each are located in \( \mathbb{R}^n \)), mollifier kernel \( k \), trained network \( f_{\theta} \), diffusion schedule \( \bar{\alpha}_t \), inverse signal-noise-ratio \( \epsilon \)
\State Sample Gaussian white noise \( \varepsilon \sim \mathcal{N}(0, I) \) 

\ForAll{coordinates \( c \in C \)}
    \State Set initial sample:
    \[
    x_T(c) \gets \int_{\mathbb{R}^n} k(c - c') \varepsilon(c') \, dc'
    \]
\EndFor

\For{\( t = T, T-1, \ldots, 1 \)}
    \State Predict mollified noise: \( \hat{\xi} \gets f_{\theta}(x_t, t) \)
    \State Estimate denoised state:
    \[
    \hat{x}_0 \gets \frac{x_t - \sqrt{1 - \bar{\alpha}_t} \cdot \hat{\xi}}{\sqrt{\bar{\alpha}_t}}
    \]
    \If{\( t > 1 \)}
        \State Sample new noise: \( \varepsilon \sim \mathcal{N}(0, I) \)
        \ForAll{coordinates \( c \in C \)}
            \State Compute mollified noise:
            \[
            \xi_{t-1}(c) \gets \int_{\mathbb{R}^n} k(c - c') \varepsilon(c') \, dc'
            \]
        \EndFor
        \State Update sample:
        \[
        x_{t-1} \gets \sqrt{\bar{\alpha}_{t-1}} \cdot \hat{x}_0 + \sqrt{1 - \bar{\alpha}_{t-1}} \cdot \xi_{t-1}
        \]
    \EndIf
\EndFor

\State Apply Wiener filtering:
\[
x_0(\omega) \gets \frac{e^{-\omega^2 t}}{e^{-2\omega^2 t} + \epsilon^2} \cdot \hat{x}_0(\omega)
\]
\State \Return \( x_0(c) \)

\end{algorithmic}
\end{algorithm}

\subsection{Network architecture}
\label{SI_Network_architecture}
The overall network architecture is designed to flexibly handle sparse and irregularly distributed observations, such as those from in-situ rain gauges, while maintaining strong representational capacity across heterogeneous data sources. As detailed in Section \ref{Network architecture}, the key distinction from a standard U-Net lies in the inclusion of multiple stacked \textit{SparseConvResBlock} modules at both the input and output stages of the network. These modules are specifically designed to process inputs with sparse spatial distributions. The input to the model consists of feature representations $x \in \mathbb{R}^{B \times L \times C}$ along with their corresponding spatial coordinates in $\mathbb{R}^{B \times L \times 2}$, where $B$ is the batch size, $L$ is the number of gauge locations, and $C$ is the number of feature channels. 
After being processed by a series of \textit{SparseConvResBlock} modules, the features retain their shape while being adapted to the sparsity of the input. These processed features are then transformed onto a coarser, structured grid, which facilitates subsequent processing using a conventional U-Net. See Supplementary Information, Listing~\ref{SI_code}, for the PyTorch implementation of the sparse-to-grid transformation.

\clearpage 
\begin{lstlisting}[language=Python, caption={Transform sparse gauge representation to a structured coarse grid.}, label={SI_code}]
from pytorch3d.ops import knn_points, knn_gather

def knn_interpolate_to_grid(x, coords, uno_coords, knn_neighbours):
    """
    Interpolates sparse features to a structured grid using KNN.

    Args:
        x: Tensor of shape (B, L, C), input features at irregular locations
        coords: Tensor of shape (B, L, 2), spatial coordinates of x
        uno_coords: Tensor of shape (y_length, 2), coordinates of target structured grid
        knn_neighbours: int, number of nearest neighbors to use
    Returns:
        Tensor of shape (B, y_length, C), interpolated features
    """

    B = x.size(0)
    target_coords = uno_coords.unsqueeze(0).repeat(B, 1, 1)  # (B, y_length, 2)

    with torch.no_grad():
        _, assign_index, neighbour_coords = knn_points(
            target_coords, coords, K=knn_neighbours, return_nn=True
        )

        # neighbour_coords: (B, y_length, K, 2)
        diff = neighbour_coords - target_coords.unsqueeze(2)
        squared_distance = (diff * diff).sum(dim=-1, keepdim=True)
        weights = 1.0 / torch.clamp(squared_distance, min=1e-15)

    neighbours = knn_gather(x, assign_index)  # (B, y_length, K, C)
    out = (neighbours * weights).sum(2) / weights.sum(2)

    return out.to(x.dtype)
    
\end{lstlisting}

\begin{figure}[htbp]
    \centering
    \includegraphics[width=\textwidth]{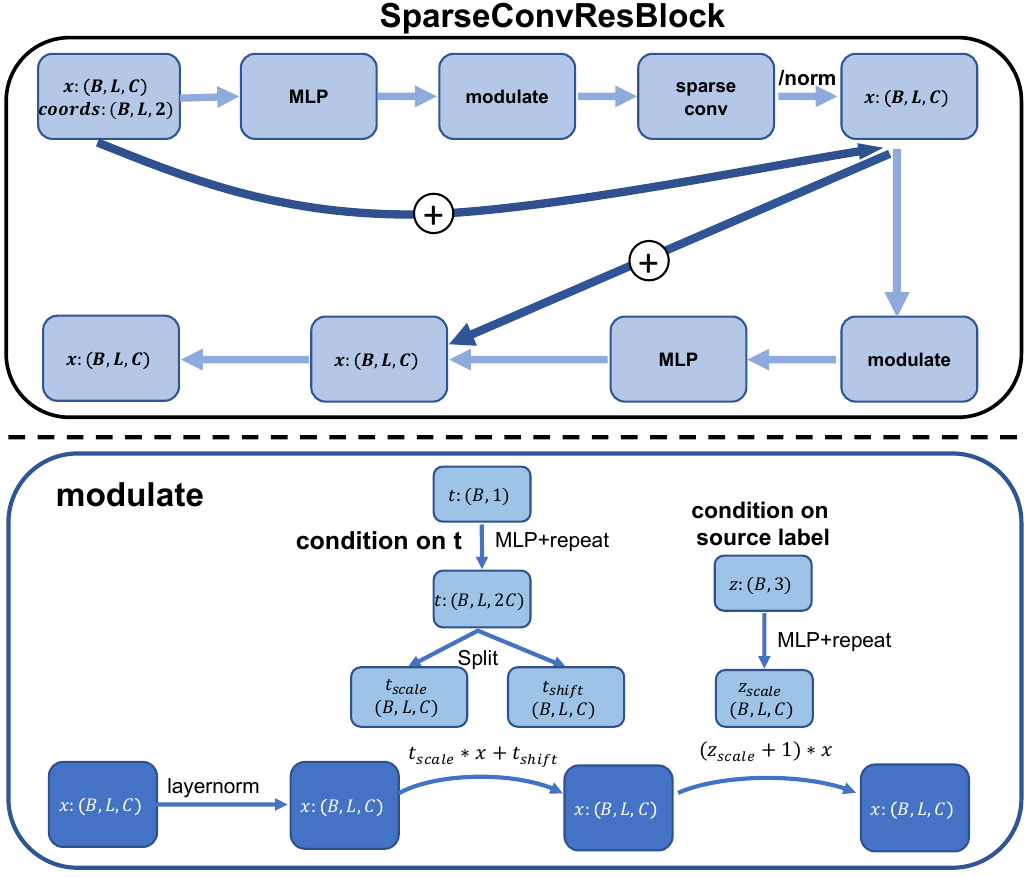}
    \caption{\textbf{Architecture of the SparseConvResBlock and its modulation mechanism.} 
    The top panel illustrates the overall structure of the \textit{SparseConvResBlock}, which processes input features $x \in \mathbb{R}^{B \times L \times C}$ and associated coordinates $\mathbb{R}^{B \times L \times 2}$ through a residual block comprising one sparse depthwise convolution. The term \textit{norm} shares the same shape as $x \in \mathbb{R}^{B \times L \times C}$ and is precomputed by convolving a unit-valued sparse tensor with fixed, non-trainable weights. The convolution kernel is initialized as a uniform averaging filter, where each weight is set to $1/K^2$ ($K$ refers the kernel size). This operation estimates local support density, ensuring numerical stability. The bottom panel shows the internal design of the \textit{modulate} module, which conditions the representation on two external variables: the diffusion timestep $t \in \mathbb{R}^{B \times 1}$ and the source label $z \in \mathbb{R}^{B \times 3}$ that denotes the dataset identity. Specifically, the source label is a one-hot vector indicating the origin of each sample, where ERA5 is represented as $(1, 0, 0)$, IMERG as $(0, 1, 0)$, and gauge observations as $(0, 0, 1)$. This encoding enables the model to learn dataset-specific feature modulations while maintaining a unified architecture across heterogeneous data sources. The timestep embedding is transformed by an MLP and split into scaling ($t_\mathrm{scale}$) and shifting ($t_\mathrm{shift}$) components, applied to the normalized input. Simultaneously, the source label contributes a scaling factor $z_\mathrm{scale}$ that further modulates the representation. This dual conditioning enables flexible control over the representation across both temporal and domain dimensions.}
    \label{fig:SI_SparseConvResBlock}
\end{figure}

Upon completion of the U-Net forward pass, the resulting features are bilinearly interpolated (using torch.nn.functional.grid\_sample function) back to the original set of irregular input coordinates. Finally, multiple \textit{SparseConvResBlock} modules are applied to further refine the outputs at target spatial locations $\mathbb{R}^{B \times L \times 2}$. The architecture of the \textit{SparseConvResBlock} module is shown in Figure~\ref{fig:SI_SparseConvResBlock}, highlighting its ability to integrate conditioning on both diffusion timestep and dataset source labels, enabling the model to operate seamlessly across multi-source inputs with varying spatial coverage.

\subsection{Rationale for the use of sparse convolution in PRIMER}
\label{proof3}

Let \( f : D \subset \mathbb{R}^2 \to \mathbb{R} \) denote a spatially continuous function defined over a bounded domain \( D \), observed only at a finite set of locations \( \mathcal{C} = \{\mathbf{c}_i\}_{i=1}^{N} \subset D \), corresponding to sparse gauge measurements. These observations form a discrete sample set \( \mathcal{S} = \{(\mathbf{c}_i, f(\mathbf{c}_i))\}_{i=1}^{N} \). We assume that \( f \) lies in a Sobolev space \( H^s(D) \), which consists of functions in \( L^2(D) \) whose weak derivatives up to order \( s \) are also square-integrable:
\[
H^s(D) = \left\{ f \in L^2(D) \mid \partial^\alpha f \in L^2(D), \; \forall |\alpha| \leq s \right\},
\]
where \( s > d/2 \) (with \( d = 2 \) in our case). This condition ensures that \( f \) is sufficiently smooth. Moreover, we assume that \( f \) is approximately band-limited in the spectral domain; that is, its Fourier transform \( \hat{f}(\boldsymbol{\omega}) \) satisfies \( \hat{f}(\boldsymbol{\omega}) \approx 0 \) for \( \|\boldsymbol{\omega}\| > \Omega \), for some cut-off frequency \( \Omega > 0 \). This implies that the function’s energy is primarily concentrated in a bounded low-frequency range. Consequently, even under sparse sampling, the dominant frequency characteristics of \( f \) are preserved, especially the low-frequency content that encodes large-scale spatial structure.

Traditional convolutional neural networks rely on regular grids, which impose translation-equivariant operations on dense Euclidean tensors. In contrast, sparse convolutional networks define a convolution operator $K_\theta$ directly on the irregular domain $\mathcal{C}$ without requiring interpolation or resampling to a dense grid. The sparse convolution operator $K_\theta$ is designed to operate on non-uniform point clouds by convolving features over local neighborhoods defined on the support $\mathcal{C}$. Given features $h$ defined at sparse locations, the sparse convolution updates features by aggregating information from neighboring points via: $(K_\theta h)(\mathbf{c}_i) = \sum_{\mathbf{c}_j \in \mathcal{N}(\mathbf{c}_i)} \kappa_\theta(\mathbf{c}_j - \mathbf{c}_i) \cdot h(\mathbf{c}_j)$. Here $\mathcal{N}(\mathbf{c}_i)$ denotes the receptive field around $\mathbf{c}_i$, and $\kappa_\theta$ are learnable kernel weights that depend on relative spatial coordinates. This formulation naturally adapts to the irregular geometry of gauge networks and preserves local spatial relationships without imposing artificial gridding.

From a mathematical standpoint, the convolution operation can be interpreted as a linear integral operator acting on the input function $f$. By the convolution theorem, applying a spatial convolution is equivalent to performing a pointwise multiplication in the frequency domain: $f * \kappa \;\longleftrightarrow\; \hat{f}(\boldsymbol{\omega}) \cdot \hat{\kappa}(\boldsymbol{\omega})$ where $\hat{f}$ and $\hat{\kappa}$ denote the Fourier transforms of $f$ and the kernel $\kappa$, respectively. This identity implies that convolutional neural networks fundamentally implement structured linear operators in the spectral domain, modulated by nonlinear activations in the spatial domain. In our setting, we consider a field $f$ that is band-limited and belongs to a Sobolev space $H^s(D)$. The band-limited assumption ensures that the energy of $\hat{f}(\boldsymbol{\omega})$ is concentrated within a compact subset of the frequency domain. Furthermore, since the gauge observations \( \mathcal{S} = \{(\mathbf{c}_i, f(\mathbf{c}_i))\}_{i=1}^{N} \) are assumed to sample \( f \) in a non-pathological  manner—that is, the sampling locations \( \{\mathbf{c}_i\} \) are well-distributed across the domain and do not systematically avoid critical regions—the low-frequency components of \( f \) are approximately preserved under such sparse sampling schemes. This stability implies that a sparse convolutional operator $K_\theta$, though defined over an irregular set $\mathcal{C}$, still implements a meaningful approximation of the frequency-domain filtering process. Therefore, the sparse convolution module can be regarded as a discretized, sampling-robust spectral operator, providing the mathematical basis for its application to be a core module in PRIMER.

\subsection{Choice of experiment parameter}
\label{Choice of experiment parameter}

PRIMER is built using the PyTorch framework~\cite{paszke2017automatic}. We summarize here the key hyper-parameters used during training and inference. Due to the computational cost associated with training PRIMER, we did not perform an extensive hyper-parameter search. Instead, all values were chosen based on empirical experience. We expect that further tuning may yield improved performance. Notably, training was intermittently paused and resumed multiple times, during which model weights were checkpointed and certain hyper-parameters were adjusted to optimize convergence. Further systematic tuning may still improve overall performance.

\begin{table}[htbp]
\centering
\caption{\textbf{Key parameters used in this study.} All values are empirically chosen without hyperparameter search.}
\renewcommand{\arraystretch}{1.3}
\begin{tabular}{p{4.2cm} p{9.8cm}}
\toprule
\textbf{Parameter} & \textbf{Description / Value} \\
\midrule
\texttt{OS} & Linux-5.10.0-34-cloud-amd64-x86\_64-with-glibc2.31 \\
\texttt{Python version} & 3.10.0 \\
\texttt{GPU count} & 2 \\
\texttt{GPU type} & NVIDIA A100-SXM4-40GB \\
\texttt{CUDA version} & 11.7 \\
\midrule
\texttt{Overall parameters} & 430,058,544 trainable parameters \\
\texttt{diffusion\_steps} & 1000 \\
\midrule
\multicolumn{2}{l}{\textbf{AdamW optimizer settings}} \\
\quad \texttt{beta1} & 0.9 (1st moment decay rate) \\
\quad \texttt{beta2} & 0.99 (2nd moment decay rate) \\
\quad \texttt{weight\_decay} & 4e-6 \\
\midrule
\multicolumn{2}{l}{\textbf{EMA (Exponential Moving Average) settings}} \\
\quad \texttt{decay} & 0.995 \\
\quad \texttt{update\_every} & every 10 batches \\
\midrule
\texttt{Batch size} & Varied between 2–6 (per GPU) due to intermittent training interruptions \\
\texttt{Learning rate} & Varied between $10^{-4}$ and $10^{-6}$ due to intermittent training interruptions \\
\bottomrule
\end{tabular}
\end{table}

\begin{figure}[htbp]
    \centering
    \includegraphics[width=\textwidth]{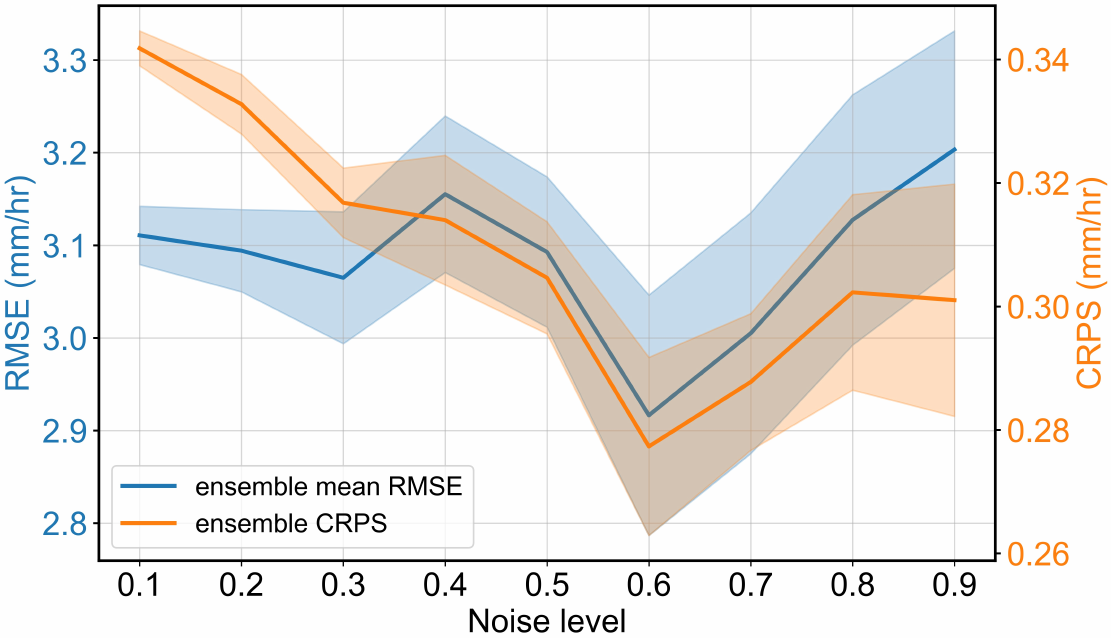}
    \caption{\textbf{Sensitivity of ensemble RMSE and CRPS to the noise level parameter \(\tau\).} 
    Evaluation is performed across a range of noise levels from 0.1 to 0.9. Ensemble members are sampled from $P_{*}(x \mid \mathcal{O}_{\mathrm{IMERG}})$, and both the ensemble-mean root mean square error (RMSE; blue) and the continuous ranked probability score (CRPS; orange) are computed over 50 repeated subsampling trials, each using 10 randomly selected members. 
    Shaded bands denote \(\pm 1\) standard deviation across repetitions. 
    Both metrics show improvement as \(\tau\) increases up to 0.6, reflecting a favorable balance between accuracy and diversity, but deteriorate beyond this point due to excessive stochasticity. 
    These results support the choice of an intermediate noise level to balance observational fidelity with generative variability.}
    \label{fig:SI_determine_alpha}
\end{figure}

\section{Details of data}

\subsection{Locations of gauges}

\begin{figure}[htbp]
    \centering
    \includegraphics[width=\textwidth]{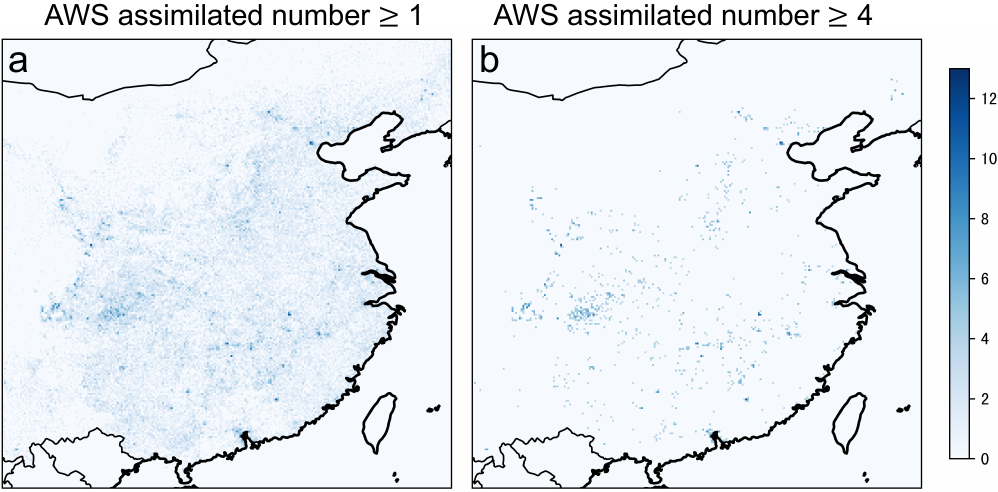}
    \caption{\textbf{Spatial distribution of assimilated AWS counts.} 
Panels show the number of automatic weather stations (AWS) assimilated into each grid point of the high spatiotemporal gauge-satellite merged precipitation analysis from~\cite{shen2014high}. a, Grid points with at least one assimilated AWS observation, representing the full set of available data used for training. b, A more stringent subset showing grid points with four or more assimilated AWS observations, used exclusively for model evaluation. This design ensures that the test regions are better constrained by in-situ observations, enabling a robust assessment of model performance.
}
    \label{fig:SI_gauges_info}
\end{figure}

\subsection{Test set}
\label{Selection of test precipitation events in 2016}
The 150 representative precipitation events used for evaluation in Fig.~\ref{fig:station_test_baseline_minus_PRIMER} and Fig.~\ref{fig:gain} were carefully selected from hourly gauge observations collected across the study domain throughout 2016. At each timestamp, approximately 1,000 stations provided precipitation measurements.
To ensure a robust and representative test dataset, we employed two complementary intensity-based selection criteria. First, we identified the 100 timestamps exhibiting the highest individual station precipitation intensities, specifically highlighting localized extreme events. Second, we selected 50 additional timestamps characterized by the highest average precipitation intensity across all stations, thus capturing widespread precipitation scenarios. These distinct yet complementary selection strategies ensure comprehensive coverage of heavy precipitation event types, enhancing the generalizability and reliability of our model evaluations. Notably, there was no overlap between these two subsets, resulting in a final, unique set of 150 precipitation events.

The test set includes the following 150 timestamps (formatted as YYYYMMDDHH): 2016070822, 2016070821, 2016071704, 2016070820, 2016061323, 2016071907, 2016081020, 2016061902, 2016072616, 2016050923, 2016040316, 2016062312, 2016070317, 2016042000, 2016071108, 2016072507, 2016082715, 2016071910, 2016071814, 2016062419, 2016080214, 2016071813, 2016070522, 2016091212, 2016050922, 2016062313, 2016070101, 2016050921, 2016071912, 2016070521, 2016061501, 2016081019, 2016081821, 2016061901, 2016072421, 2016082506, 2016061322, 2016071915, 2016071709, 2016071914, 2016070919, 2016060119, 2016071909, 2016091420, 2016060120, 2016071901, 2016071107, 2016071904, 2016072611, 2016070716, 2016062310, 2016062200, 2016080216, 2016090916, 2016060611, 2016071900, 2016080307, 2016080923, 2016082714, 2016070819, 2016081008, 2016062302, 2016080200, 2016061214, 2016060612, 2016081021, 2016062304, 2016050607, 2016060414, 2016070102, 2016091112, 2016062201, 2016071908, 2016072121, 2016071913, 2016070405, 2016071521, 2016082516, 2016070323, 2016062718, 2016072612, 2016061104, 2016070400, 2016070823, 2016061913, 2016071312, 2016052005, 2016082517, 2016071702, 2016092808, 2016072122, 2016091600, 2016080509, 2016061903, 2016061820, 2016062005, 2016051423, 2016052002, 2016070519, 2016062802, 2016071406, 2016102019, 2016071407, 2016102018, 2016102021, 2016102020, 2016052218, 2016052104, 2016071405, 2016071408, 2016102023, 2016102022, 2016052217, 2016052219, 2016071409, 2016102100, 2016061108, 2016102117, 2016052220, 2016110719, 2016102015, 2016102118, 2016061109, 2016052211, 2016071106, 2016012817, 2016110720, 2016102121, 2016061107, 2016012816, 2016052103, 2016012803, 2016102120, 2016112305, 2016013112, 2016052023, 2016061111, 2016052100, 2016110823, 2016012815, 2016102122, 2016012813, 2016102119, 2016100622, 2016071410, 2016112220, 2016102103, 2016102102, 2016102101, 2016102116.


\clearpage
\newpage

\section{Additional results}

\begin{figure}[htbp]
    \centering
\includegraphics[width=0.9\textwidth,page=1]{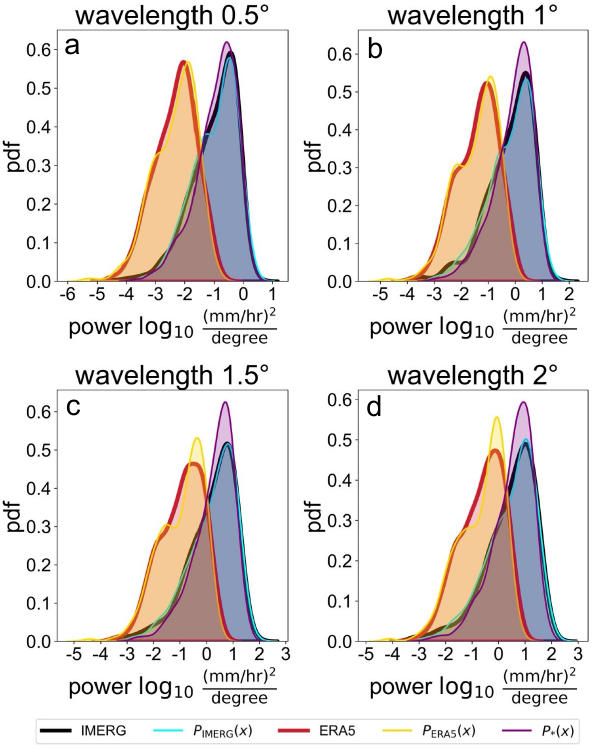} 
	\caption{\textbf{PDF of RAPSD between learned priors and reference datasets.} (\textbf{a}-\textbf{d}) shows the PDF of power at 0.5°, 1°, 1.5°, 2° wavelength respectively. 
All statistics are derived from 1,000 randomly sampled realizations of precipitation fields.
}
    \label{SI:SI_pdf_RAPSD}
\end{figure}

\begin{figure}[htbp]
    \centering
    \includegraphics[width=0.8\textwidth,page=1]{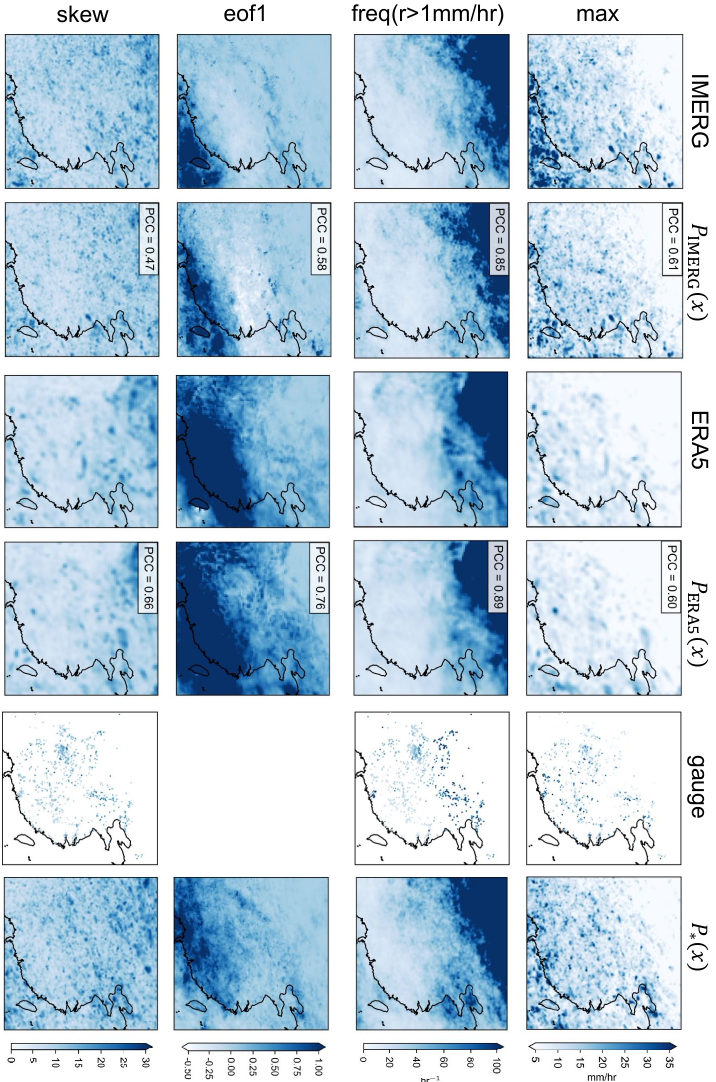} 
    \caption{\textbf{Climatological structure comparison between learned priors and reference datasets.}
For clarity, the figure has been rotated 90° clockwise.
Each row presents a distinct climatological statistic: \textbf{top to bottom}, spatial distribution of maximum precipitation rate, frequency of precipitation events ($>1$ mm/hr), the leading empirical orthogonal function (EOF1), and skewness. Each column corresponds to a different data source: IMERG, unconditional samples from $P_{\mathrm{IMERG}}(x)$, ERA5, unconditional samples from $P_{\mathrm{ERA5}}(x)$, gauge observations, and samples from the final updated prior $P_{*}(x)$. Panels associated with $P_{\mathrm{IMERG}}(x)$ and $P_{\mathrm{ERA5}}(x)$ display Pearson correlation coefficients (PCCs) with their respective reference datasets (IMERG and ERA5), highlighting structural agreement. Colorbars denote the units of each diagnostic. All statistics are derived from 1,000 randomly sampled realizations of precipitation fields.
}
    \label{SI:SI_eof_skew}
\end{figure}

\begin{figure}[htbp]
\centering
\includegraphics[width=1\textwidth,page=1]{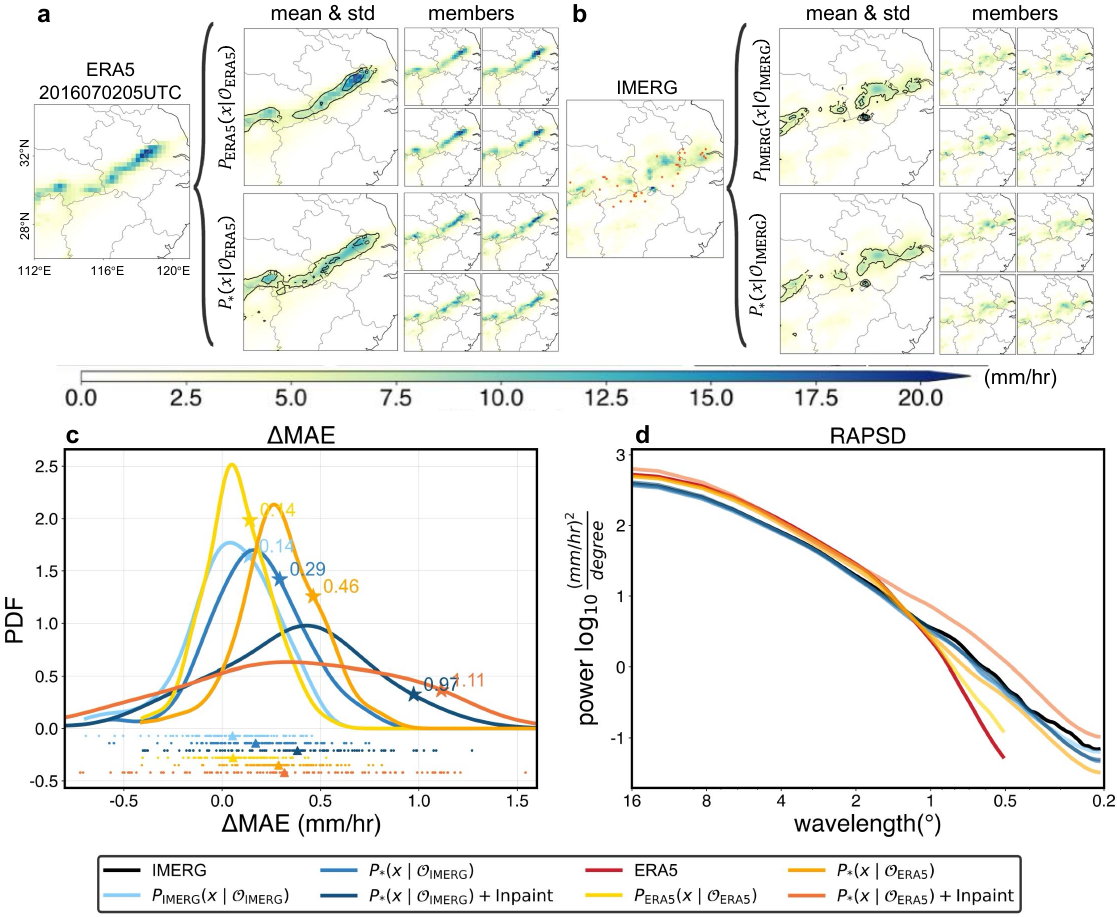}
\caption{\textbf{Case study of a Meiyu event.} 
This figure complements Fig.~\ref{fig:meiyu_hubei} by illustrating additional results sampled from alternative posterior distributions.
\textbf{a,b}, Posterior samples based on ERA5 and IMERG as conditional input, respectively. We show the original precipitation field, the posterior mean and standard deviation, and four representative ensemble members. 
\textbf{c}, Probability density functions (PDFs) of changes in mean absolute error ($\Delta$MAE) relative to original ERA5 or IMERG, with positive values indicating improved accuracy after bias correction. 
\textbf{d}, Radially averaged power spectral density (RAPSD) curves demonstrate that prior $P_{*}(x)$ effectively compensates for the underestimation of high-frequency spectral power in ERA5, thereby enhancing spatial structure realism. Note that we first interpolate ERA5 and samples from $P_{\mathrm{ERA5}}(x \mid \mathcal{O}_{\mathrm{ERA5}})$ to 0.1 degree before RAPSD caculation. 
Overall, this case highlights the flexibility of PRIMER in performing posterior sampling using diverse precipitation priors, including those derived from reanalysis, satellite. Among them, the prior $P_{*}(x)$ yields the most accurate reconstructions, underscoring the value of incorporating sparse yet reliable gauge observations for fine-tuning probabilistic models.
}

\label{SI:SI_meiyuhubei}
\end{figure}

\begin{figure}[htbp]
\centering
\includegraphics[width=1\textwidth,page=1]{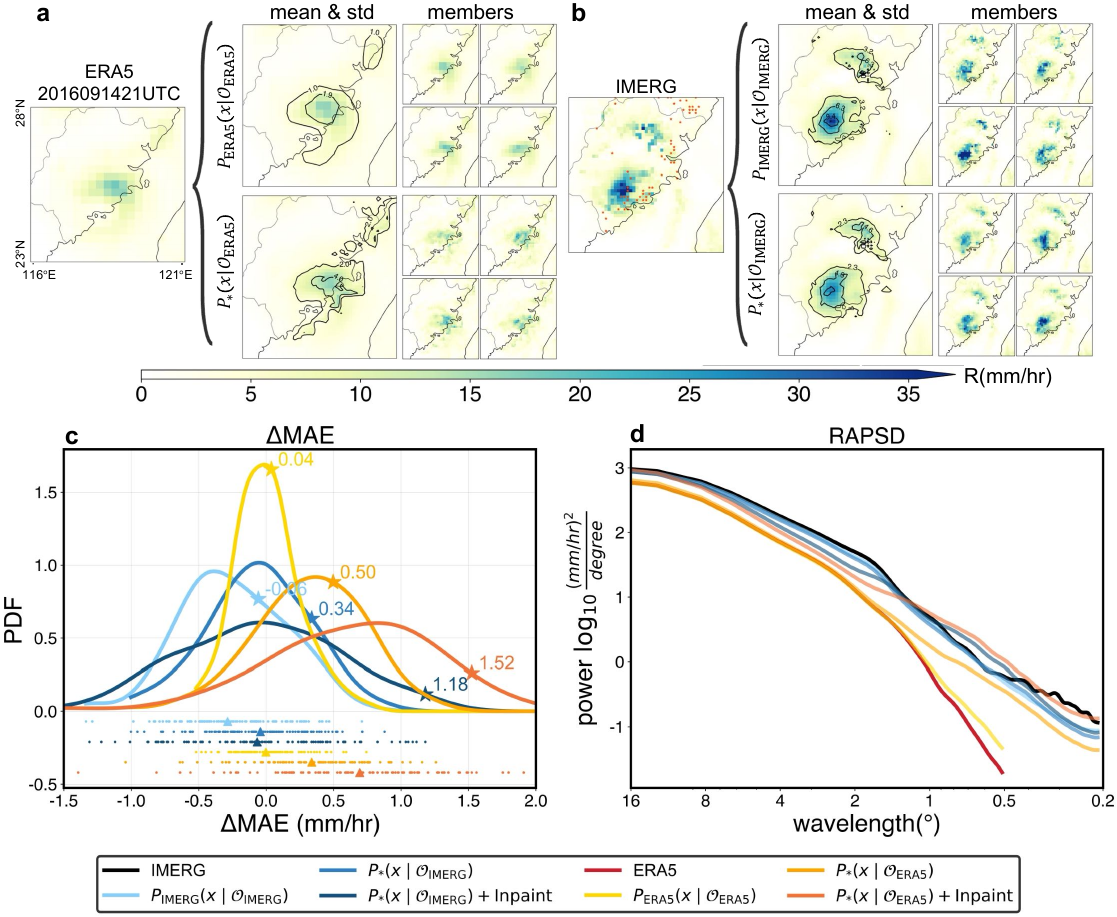}
\caption{\textbf{Case study of Typhoon Meranti (2016) precipitation event.} Typhoon Meranti, one of the most intense tropical cyclones recorded globally in 2016, made landfall in southeastern China in mid-September, causing widespread flooding and infrastructure damage. This figure is similar to Fig.~\ref{SI:SI_meiyuhubei}.}

\label{SI:SI_Meranti}
\end{figure}

\begin{figure}[htbp]
\centering
\includegraphics[width=1\textwidth,page=1]{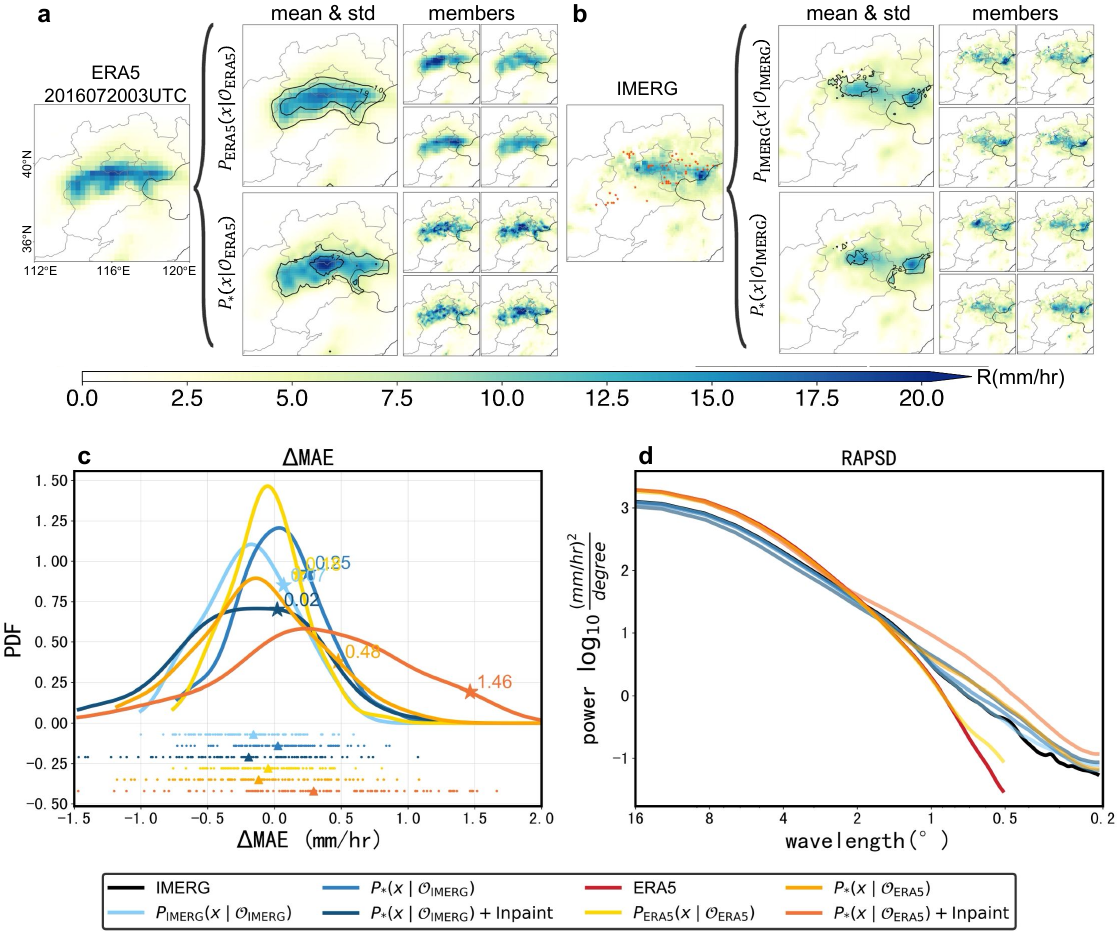}
\caption{\textbf{Case study of an extreme precipitation event near Beijing.} On 20 July 2016, an extratropical cyclone developed over North China, bringing prolonged and intense precipitation to the Beijing-Tianjin-Hebei region. This event, known as the ``7·20" rainstorm. This figure is similar to Fig.~\ref{SI:SI_meiyuhubei}.}
\label{SI:SI_720event}
\end{figure}

\begin{figure}[htbp]
\centering
\includegraphics[width=0.9\textwidth]{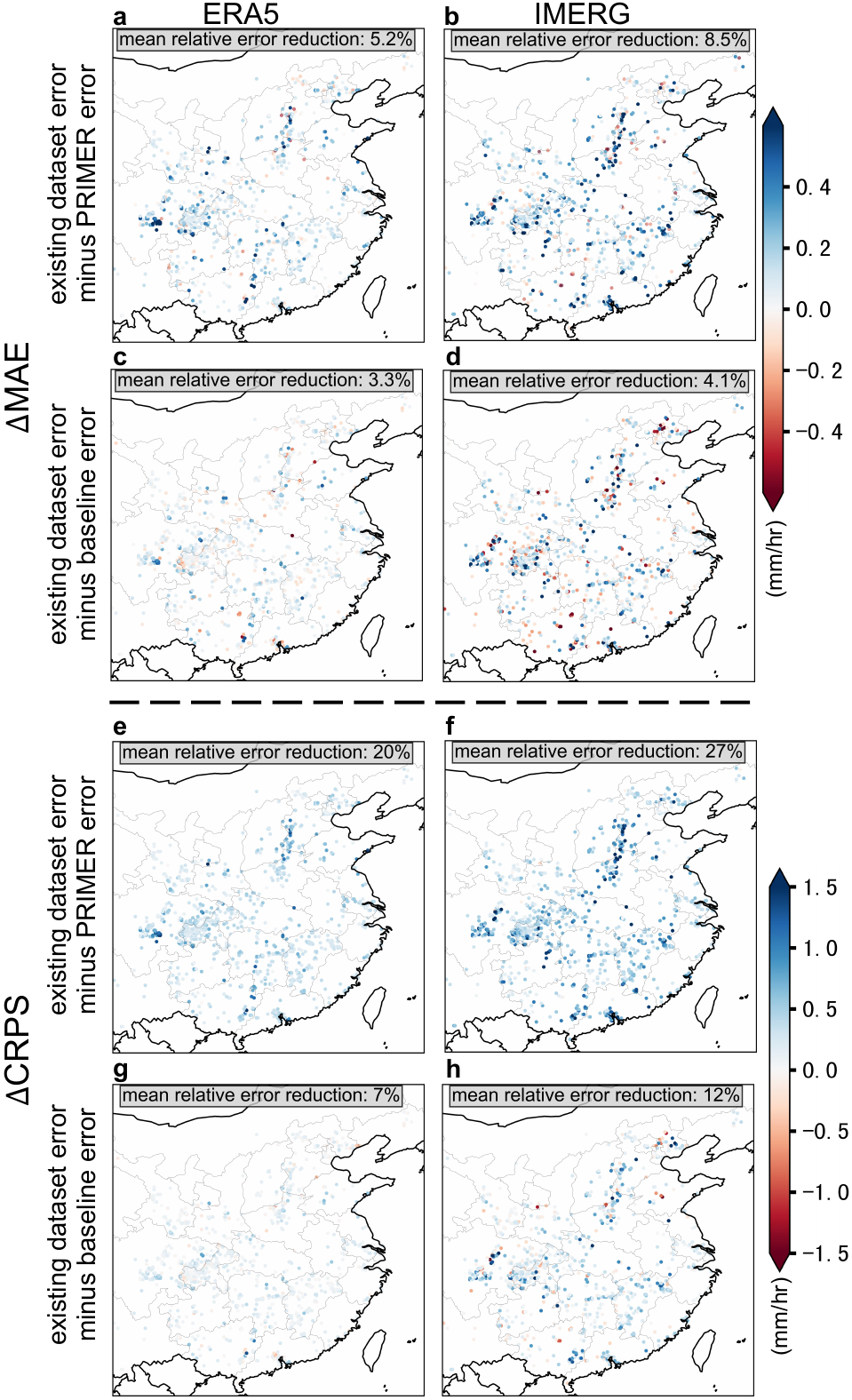}

\caption{\textbf{Spatial distributions of $\Delta \mathrm{MAE}$ and $\Delta \mathrm{CRPS}$.} 
As in Fig.~\ref{fig:station_test_baseline_minus_PRIMER}, but showing the reduction in MAE and CRPS after bias correction using PRIMER (prior $P_{*}(x)$) and the baseline priors ($P_{\mathrm{ERA5}}(x)$ and $P_{\mathrm{IMERG}}(x)$), applied separately to ERA5 and IMERG. 
The evaluation is based on 150 precipitation events that occurred in 2016. 
Overall, PRIMER outperforms the baseline method, as evidenced by larger mean relative error reductions (annotated in the top-left corner of each panel).
}
\label{SI:SI_station_test}
\end{figure}

\begin{figure}[htbp]
\centering
\includegraphics[width=1\textwidth,page=1]{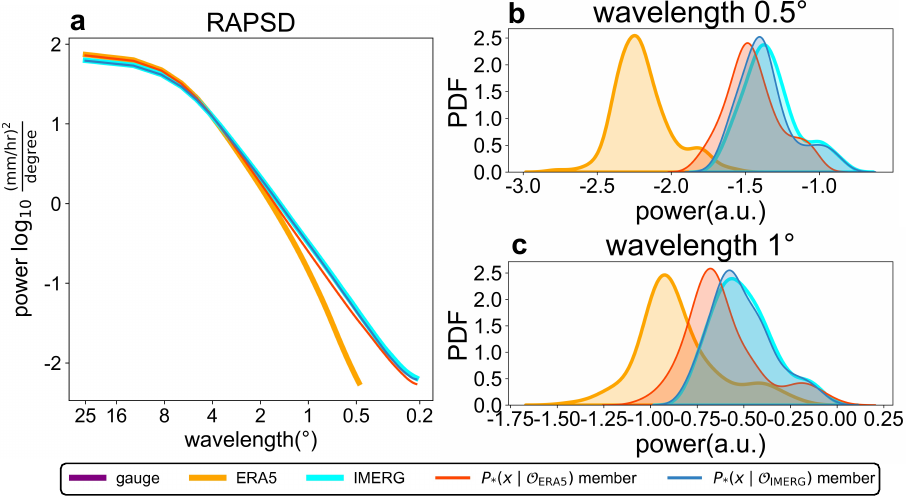}

\caption{\textbf{Enhancement of spatial variability in test datasets.} 
Due to the lack of power spectral references based on gauge observations, IMERG (0.1° resolution) is used as a proxy for evaluating fine-scale precipitation features. 
\textbf{a}, Radially averaged power spectral density (RAPSD) of log-transformed precipitation intensity, showing that PRIMER effectively restores high-frequency variability absent in the original ERA5 data. 
For consistency, ERA5 data are interpolated to a 0.1° grid prior to RAPSD computation. 
\textbf{b}, Probability density functions (PDFs) of spectral power at wavelengths of 0.5° (top) and 1° (bottom). 
While ERA5 (0.25° resolution) underrepresents spectral power at these smaller scales, samples that are generated from posteriors $P_{*}(x \mid \mathcal{O}_{\mathrm{ERA5}})$ shift the distribution toward higher power, indicating improved representation of fine-scale structure.
}

\label{SI:SI_150events_RAPSD}
\end{figure}

\begin{figure}[htbp]
\centering
\includegraphics[width=1\textwidth,page=1]{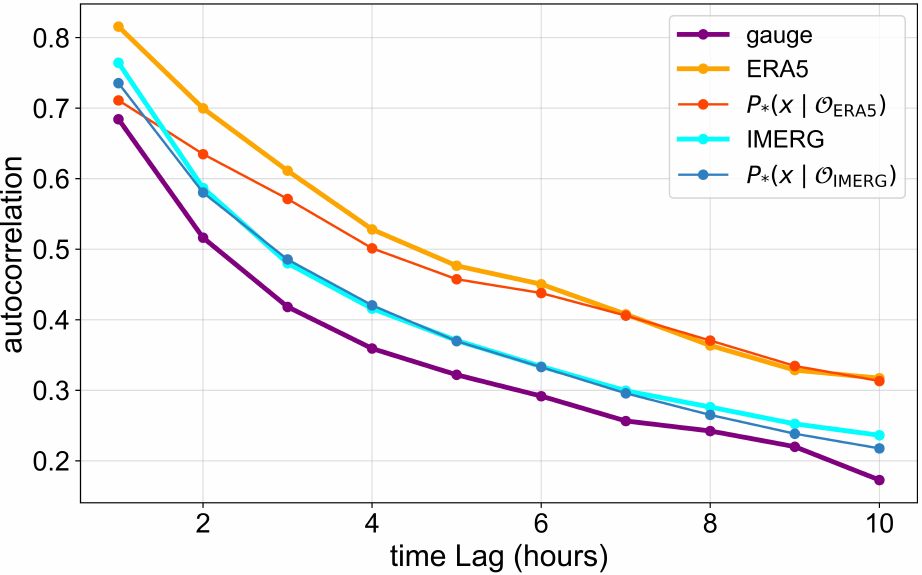}

\caption{\textbf{Temporal correlations in test datasets.} As a supplementary analysis to Fig.~\ref{fig:gain}, temporal correlations are assessed by computing the autocorrelation with a lag of up to 10 hours. Correlations are computed exclusively at gauge locations and averaged over all paired precipitation events from a subset of 2016, for gauge observations, ERA5, $P_{*}(x \mid \mathcal{O}_{\mathrm{ERA5}})$, IMERG, and $P_{*}(x \mid \mathcal{O}_{\mathrm{IMERG}})$. Results show that applying PRIMER to ERA5 and IMERG preserves the intrinsic temporal dynamics, as evidenced by comparable autocorrelation structures before and after correction. Notably, original ERA5 and IMERG exhibit higher temporal correlations than gauge observations, reflecting artificial persistence introduced probably by numerical model, data assimilation and satellite retrieval processes. After PRIMER mollification, the temporal correlations of $P_{*}(x \mid \mathcal{O}_{\mathrm{ERA5}})$ and $P_{*}(x \mid \mathcal{O}_{\mathrm{IMERG}})$ decrease and become closer to those observed in the gauge observations, indicating improved physical realism.}

\label{SI:SI_time_correlation}
\end{figure}

\begin{figure}[htbp]
\centering
\includegraphics[width=0.9\textwidth]{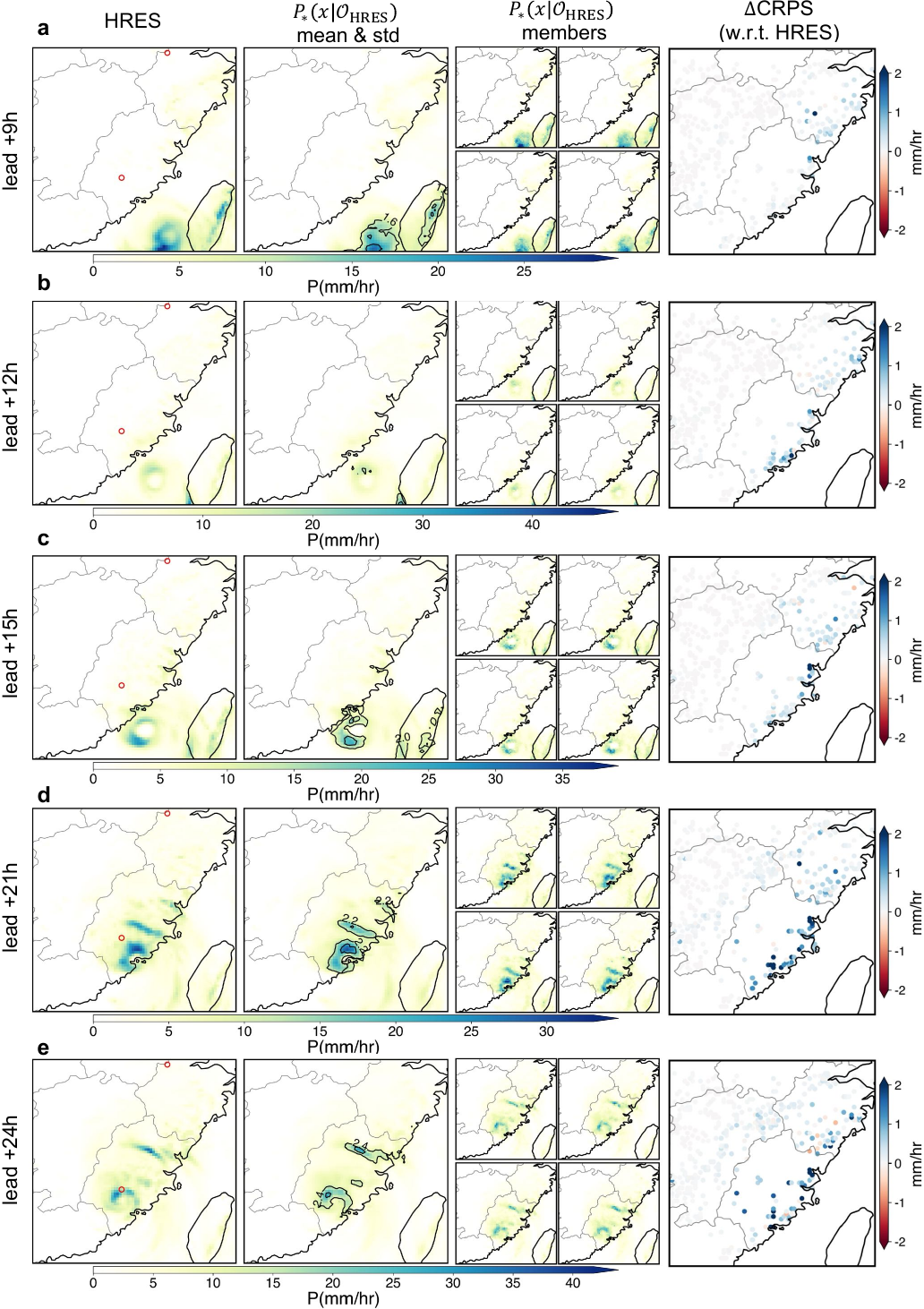}
\caption{\textbf{Zero-shot enhancement for operational forecasts.} Similar to Fig.~\ref{fig:HRES}, but for other lead times with a comprehensive illustration.}
\label{SI:SI_HRES}
\end{figure}

\begin{figure}[htbp]
\ContinuedFloat
\centering
\includegraphics[width=0.9\textwidth]{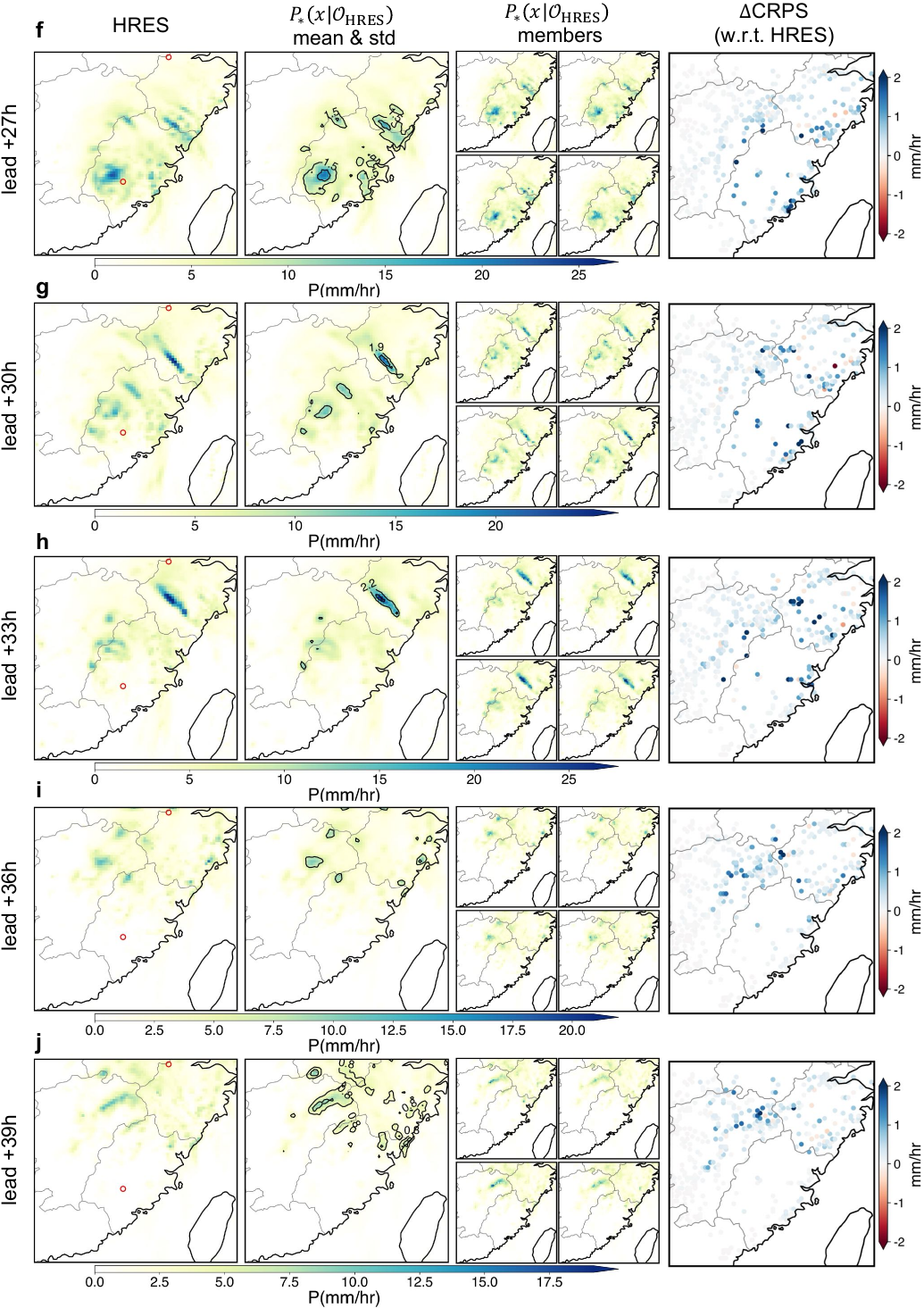}
\caption{\textbf{Continued from previous page.}}
\end{figure}

\begin{figure}[htbp]
\centering
\includegraphics[width=1\textwidth]{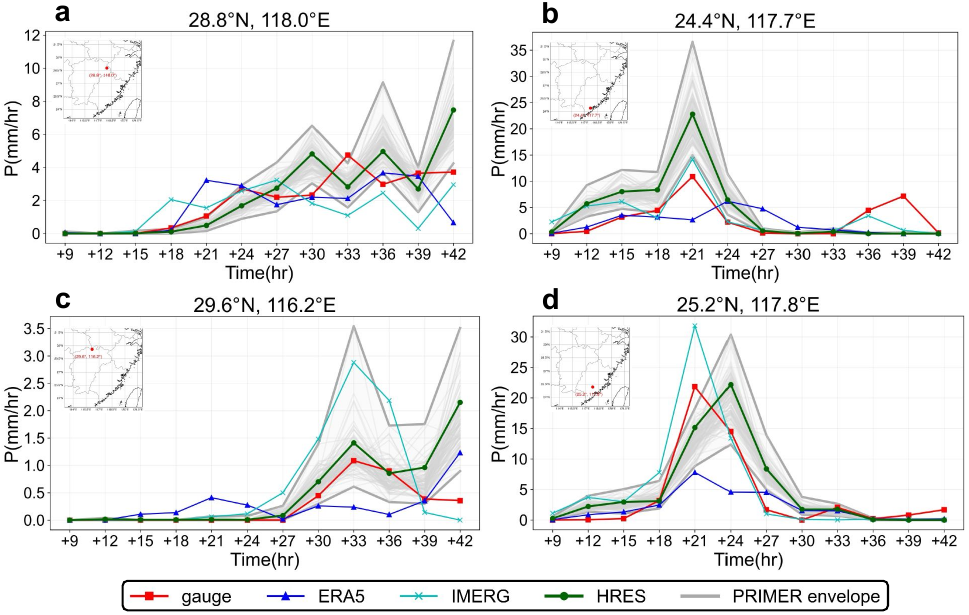}
\caption{\textbf{Zero-shot enhancement for operational forecasts.} Similar to Fig.~\ref{fig:HRES}, but for precipitation time series at other representative gauge stations; gray envelope denotes the spread across 100 ensemble members.}
\label{SI:SI_HRES_representative_station}
\end{figure}

\end{appendices}


\clearpage
\newpage

\bibliography{main}

\end{document}